\documentclass[journal]{IEEEtran}
\ifCLASSINFOpdf
\else
\fi
\usepackage{amsmath}
\usepackage{amsfonts}
\usepackage{amssymb}
 \usepackage{array}
 \usepackage{graphicx}
 \usepackage{epstopdf}
 \usepackage{balance}
\usepackage{colortbl}
\usepackage{graphicx}
\usepackage{epstopdf}
\usepackage{algorithm}
\usepackage{booktabs}
\usepackage{bm}
\usepackage{array}

\usepackage{cite}
\usepackage{multirow}
\usepackage{subfig}
\usepackage{float}

\ifCLASSOPTIONcaptionsoff
  \usepackage[nomarkers]{endfloat}
 \let\MYoriglatexcaption\caption
 \renewcommand{\caption}[2][\relax]{\MYoriglatexcaption[#2]{#2}}
\fi
\begin{document}
%
\title{Target matching based generative model for speech enhancement}
%
%

\author{Taihui~Wang, Rilin~Chen, Tong~Lei,~\IEEEmembership{Student Member,~IEEE,} Andong~Li,~\IEEEmembership{Member,~IEEE,}, Jinzheng~Zhao,
        Meng~Yu,~\IEEEmembership{Member,~IEEE,} and Dong Yu,~\IEEEmembership{Fellow,~IEEE}

\thanks{Manuscript received -- --, 2025; revised -- --, 2025.}
\thanks{Taihui Wang, Rilin Chen, and Jinzheng Zhao are with Tencent  AI Lab, Beijing 100193, China, and also with Tencent Multimodal Models Department, Beijing 100193, China. (Corresponding author: Rilin Chen)}
\thanks{Lei Tong is with Tencent AI Lab, Shenzhen,  China.}
\thanks{Andong Li is with the Key Laboratory of Noise and Vibration Research, Institute of Acoustics, Chinese Academy of Sciences, Beijing, 100190, China.}
\thanks{Meng Yu and Dong Yu are with Tencent AI Lab, Bellevue, WA, USA.}
\thanks{ \textit{Email: \{taihuiwang, rilinchen, coffeyzhao, fayelei\}@tencent.com, liandong@mail.ioa.ac.cn,  \{raymondmyu, dyu\}@global.tencent.com}. }
}

%
%

\markboth{IEEE TRANSACTIONS ON AUDIO, SPEECH, AND LANGUAGE PROCESSING}%
{Shell \MakeLowercase{\textit{et al.}}: Bare Demo of IEEEtran.cls for IEEE Journals}
%



\maketitle

\begin{abstract}
The design of mean and variance schedules for the perturbed signal is a fundamental challenge in generative models. While score-based and Schr\"odinger  bridge-based models require careful selection of the stochastic differential equation to derive the corresponding schedules, flow-based models address this issue via vector field matching.
However, this strategy often leads to  hallucination artifacts and inefficient training and inference processes due to the potential inclusion of stochastic components in the vector field. Additionally, the widely adopted diffusion backbone, NCSN++, suffers from high computational complexity. 
To overcome these limitations, we propose a novel target-based generative framework that enhances both the flexibility of mean/variance schedule design and the efficiency of training and inference processes. 
Specifically, we eliminate the stochastic components in the training loss by reformulating the generative speech enhancement task as a target signal estimation problem, which therefore leads to more stable and efficient training and inference processes. In addition, we employ a logistic mean schedule and a bridge variance schedule, which yield a more favorable signal-to-noise ratio trajectory compared to several widely used schedules and thus leads to a more efficient perturbation strategy.
 Furthermore, we propose a new diffusion backbone for audio, which significantly improves the efficiency over NCSN++ by explicitly modeling long-term frame correlations and cross-band dependencies. 
 The key advantages of the proposed framework are fourfold: (1) it supports arbitrary mean and variance schedules, offering greater design flexibility; (2) it ensures efficient training and inference by optimizing a deterministic target objective; (3) it leverages the signal-level loss to improve the target estimation accuracy; and (4) it uses a new dual-path spectro-temporal diffusion backbone for audio, which is efficient and further supports scalable model configurations. Experimental results demonstrate the superiority of the proposed approach in speech enhancement performance, model size and computational efficiency.
\end{abstract}


%
\IEEEpeerreviewmaketitle

\section{Introduction}
%
%
%
%
\IEEEPARstart{T}{he} degradation of speech signals in noisy environments persists as a pervasive challenge, not only impairing human auditory perception but also compromising the performance of machine-driven speech processing applications, such as automatic speech recognition \cite{arakiImpactResidualNoise2023} and speaker verification \cite{raoTargetSpeakerExtraction2019}.  Consequently, the development of advanced speech enhancement techniques targeting noise suppression \cite{SpeechEnhancementTheory} have become a fundamental requirement for enabling clear speech communication and ensuring the reliability of downstream systems.

Traditional statistical-model-based speech enhancement approaches typically rely on modeling the statistical properties of speech signals \cite{benestySpeechEnhancement2006}, yet their performance remains fundamentally constrained due to the inherent limitations of statistical models in accurately characterizing real-world acoustic environments. In contrast, deep learning-based speech enhancement frameworks have demonstrated superior performance by adopting data-driven methodologies to directly learn either the spectral mapping or time-frequency masking functions between noisy observations and clean speech signals \cite{wangSupervisedSpeechSeparation2018a,kuangThreestageHybridNeural2023,sun2024smru}. 
Nevertheless, discriminative approaches frequently introduce perceptually unnatural artifacts, primarily due to their inability to accurately model the underlying statistical distribution of speech signals. Moreover, these methods often exhibit poor generalization performance, as they tend to overfit to specific training data patterns and struggle to adapt to unseen acoustic conditions. 
These limitations have motivated the adoption of generative modeling paradigms that explicitly learn the data distribution of clean speech signals, such as the generative adversarial networks \cite{pascualSEGANSpeechEnhancement2017}, variational autoencoders \cite{fangVariationalAutoencoderSpeech2021}, and denoising diffusion probabilistic models \cite{luConditionalDiffusionProbabilistic2022}.

Within these generative speech enhancement methods, the score-matching \cite{songSCOREBASEDGENERATIVEMODELING2021a} based diffusion models have emerged as a particularly promising framework \cite{welkerSpeechEnhancementScoreBased2022, richterSpeechEnhancementDereverberation2023, gonzalezInvestigatingDesignSpace2024}. These methods transform the noisy prior distribution to the clean speech distribution step by step based on the learned time-conditioned score model, where a neural network is trained to match the time-varying score function given the time-conditioned perturbed signal. It can be proved that the perturbed signal follows the Gaussian distribution with the time-varying mean and variance \cite{sarkkaAppliedStochasticDifferential2019}, which is called mean and variance schedules. The closed-form solutions of the mean and variance schedules are derived based on the predefined stochastic differential equation (SDE). Several different SDEs have been utilized in order to design effective mean and variance schedules, such as the Ornstein-Uhlenbeck process with the variance exploding (OUVE) \cite{welkerSpeechEnhancementScoreBased2022, richterSpeechEnhancementDereverberation2023} or  variance preserving (OUVP) \cite{gonzalezInvestigatingDesignSpace2024}, and the Brownian bridge process with exponential diffusion (BBED) \cite{laySingleFewStepDiffusion2024a}. It can be concluded from these works that the mean and variance schedule have a significant impact on the performance of the score-matching based diffusion models \cite{gonzalezInvestigatingDesignSpace2024}. The reason is that the mean and variance schedules determine the effectiveness and efficiency of adding Gaussian noise when training the score model.
This motivates us to explore more optimized mean and variance scheduling strategies. However, we note that the design of the mean and variance schedule in diffusion models is very difficult, because they cannot be defined directly but be solved from the predefined SDE.

To address this problem, flow-matching (FM) based generative methods \cite{lipmanFlowMatchingGenerative2023, tongImprovingGeneralizingFlowbased2024} model the distribution transport problem using an ordinary differential equation (ODE) governed by a vector field, rather than relying on the score function within an SDE. Unlike the score which is calculated using the predefined SDE, the vector field can be directly derived from predefined mean and variance schedules. This flexibility enables us to design customized mean and variance schedules for generating more efficient perturbed signals.
Several mean and variance schedules have been proposed for the speech enhancement task \cite{korostikModifyingFlowMatching2025, leeFlowSEFlowMatchingbased2025, jungFlowAVSEEfficientAudioVisual2024}. In \cite{leeFlowSEFlowMatchingbased2025}, the linear and time-varying mean and variance schedules are used  to design a speech enhancement method. 
However, we observe that employing a time-variant variance schedule leads to a stochastic vector field, which can not be learned efficiently. Consequently, the inference process becomes inefficient, requiring a large number of steps. Additionally, this stochastic vector field is likely to introduce hallucination artifacts in low frequency regions under low signal-to-noise ratio (SNR) conditions, where speech-spectrum-shaped noise devoid of actual linguistic content is generated. 

As discussed above, a neural network must be used to parameterize the score function in the score-matching based methods or the vector field in the flow-matching based methods, and hence designing an efficient neural network architecture is important. The network architecture design of generative models is inherently more complex than that of discriminative models, as it necessitates the processing of multiple timesteps. The widely used backbone in audio tasks is the Noise Conditional Score Network++ (NCSN++), which was first proposed in \cite{songSCOREBASEDGENERATIVEMODELING2021a} for the image generation task and then has been widely used in speech-related tasks. NCSN++ is well-suited for image signals because its 2D convolutional operations inherently capture spatial locality and translation invariance, while its multi-scale noise scheduling mechanism effectively models structured dependencies among pixels. However, we argue that this architecture is not directly applicable to audio signals, as time-frequency representations require decoupled processing of temporal and frequency dimensions. Specifically, the temporal dimension demands modeling long-range dependencies (e.g., contextual coherence in speech), whereas the frequency dimension must resolve harmonic structures and energy distributions. In contrast, 2D convolutions in image processing struggle to explicitly disentangle these time-frequency characteristics, potentially leading to phase misalignment or over-smoothed spectra in audio reconstruction.

To address these fundamental limitations, we propose a target-matching (TM) based generative framework that reformulates the distribution transport problem. Unlike flow-matching methods requiring estimation of stochastic vector fields, our approach directly predicts the deterministic target signal through a neural network. This design eliminates stochastic components in the optimization objective, consequently enabling faster training convergence, requiring fewer inference steps for stable output, and reducing the hallucination artifacts over conventional flow-matching methods. 
Morever, the target signal's physical interpretability enables us to use the signal-level loss to enhance the estimation accuracy, which cannot be achieved in score or flow matching methods, because the score and flow has no clean physical meaning in the signal level. 
The proposed framework also supports the flexible design of mean and variance schedules, and hence we adopt more complex logistic mean and bridge variance schedules to generate the perturbed signal, which yilds a more efficient noise scheduling process compared to the widely used linear mean and variance schedules. 
Moreover, we propose a new diffusion backbone for audio (DBA) signal processing, which uses the dual-path architecture to model long-term frame correlations and cross-band dependencies. Compared to the widely used NCSN++, the proposed DBA is more efficient and supports scalable model configurations.
Ablation studies on the noise schedules, estimation targets and diffusion backbones show the effectiveness of the proposed framework, and the comparison results show the superiority of the proposed generative methods over other generative methods.

\section{Problem Formulation}
In the short-time Fourier transform (STFT) domain, let $x_1(f,k)$ denote the observed single-channel noisy speech signal by the microphone, where $f \in \{1, . . . , F \}$ denotes the frequency bin index, and $k \in \{1, . . . , K \}$ denotes the frame index. The noisy speech signal can be represented as 
\begin{equation}\label{e1}
	x_1(f,k)=x_0(f,k)+n(f,k),
\end{equation}
where $x_0(f,k)$  and $n(f,k)$ denote the clean speech and noise signals, respectively. In the following discussion, we model all time-frequency bins independently, and thus the index $f$ and $k$ are omitted for conciseness unless stated otherwise.

The target of speech enhancement is to estimate the clean speech signal from a noisy speech signal. Generative models are introduced and treat the speech enhancement task as a distribution transport problem, i.e., from the noisy speech distribution $q_1 (x_1 )$ to the clean speech distribution $q_0 (x_0 )$. The distribution transport problem can be modeled using the differential equation (DE), and the corresponding inverse DE can be used to recover the clean speech distribution given the noisy speech distribution \cite{richterSpeechEnhancementDereverberation2023, jukicSchrodingerBridgeGenerative2024, leeFlowSEFlowMatchingbased2025}. The inverse DE often cannot be tractable and hence the parametric neural network is employed to estimate the non-tractable component in the inverse DE. Once the parametric neural network is trained, the clean speech distribution can be obtained using the inverse DE with this estimator from the noisy speech distribution. 

Therefore, the core problem in the generative speech enhancement is the design of the DE and the parametric neural network.

\section{Related Works}
In this section, we provide a concise introduction to two classes of generative speech enhancement methods: score-matching (based on SDE) and flow-matching (based on ODE) approaches.
We review their respective methodologies and highlight their current limitations.
Additionally, we introduce the NCSN++ backbone commonly adopted by these methods.
\subsection{Score-matching based Method}
In the score-matching based speech enhancement \cite{richterSpeechEnhancementDereverberation2023}, an SDE is formulated to characterize the distribution transport from clean speech $x_0$ to noisy speech $x_1$ as
\begin{equation}\label{eq:sde}
x_t=f(x_t,t,x_1 )dt+g(t)dw,
\end{equation}
where $x_t$ denotes the perturbed signal at timestep $t$, $w$ denotes a standard Wiener process, and $f(x_t,t,x_1) $ and  $g(t)$ denote the drift and diffusion coefficients, respectively. The clean distribution is gradually converged into the noisy distribution through the SDE. Following Anderson and Song \cite{andersonReversetimeDiffusionEquation1982}, Eq. (\ref{eq:sde}) admits the following reverse-time SDE 
\begin{equation}\label{eq:reverse sde}
dx_t=[ f(x_t,t,x_1)-g^2 (t) \nabla_{x_t}  \log  p_t (x_t \mid x_1 ) ] dt+g(t)d\bar{w},
\end{equation}
where $\overline{w}$ denotes the backward standard Wiener process, and $
\nabla_{x_t} \log p_t(x_t|x_1)$ is the gradient of the log-probability density function, i.e., the score function. It is shown that the current state $x_t$ follows the Gaussian distribution with the mean schedule  $\mu_t(x_0,x_1 )$  and the variance schedule $\sigma^2 (t)$, which leads to the score function $
\nabla_{x_t} \log p_t(x_t|x_1) = -(x_t - \mu_t(x_0, x_1))/\sigma^2(t)$.  The closed-form solutions for mean and variance schedules can be obtained based on Eqs. (5.50, 5.53) in Sarkka $\&$ Solin \cite{sarkkaAppliedStochasticDifferential2019} given the predefined SDE. The Ornstein-Uhlenbeck SDE is adopted in \cite{welkerSpeechEnhancementScoreBased2022,richterSpeechEnhancementDereverberation2023} and leads to the following mean schedule
\begin{equation}\label{eq:ouve mean schedule}
	{\mu}_t\left({x}_{0}, x_1\right)=\mathrm{e}^{-\gamma t} x_{0}+\left(1-\mathrm{e}^{-\gamma t}\right) x_1,
\end{equation}
where $\gamma$ is a constant to control the transition.
This solution is related to the unknown clean signal $x_0$,  and thus a neural network is used as the estimator of the parametric score function $s_{\theta}(x_t, x_1, t)$, which is trained using the score matching loss
\begin{equation}\label{e4}
\mathcal{L}_{sm}(\theta) = \mathbb{E}_{x_t|x_1,t,x_0} \left[\left| s_\theta(x_t,x_1,t) - 
\nabla_{x_t} \log p_t(x_t|x_1) \right|^2\right].
\end{equation}
Once the score network $s_{\theta}(x_t, x_1, t)$ is trained, we can iteratively obtain the clean speech signal $x_0$ from $x_1$ based on the reverse SDE (\ref{eq:reverse sde}). In this process, the score network progressively refines the estimate by taking the time-varying perturbed signal $x_t$ as input.

The perturbed signals play a fundamental role in score-matching methods, as they govern the temporal evolution of the marginal distribution. These signals are generated during the training process according to the predefined SDE, which determines their mean and variance schedules.
While existing scheduling approaches (e.g., OUVE, BBED, SBVE, SBVP) provide formal theoretical guarantees, we identify significant challenges in developing novel scheduling schemes due to fundamental constraints. The key limitation arises from the inherent coupling between schedule design and SDE formulation, because mean and variance trajectories have to strictly conform to the predefined SDE structure.
This architectural constraint substantially limits the design flexibility of score-matching methods, potentially restricting their adaptability to diverse generative tasks and hindering further performance improvements.

\subsection{Flow-matching based Method}
The flow matching is introduced to solve this problem above. In these methods, the transformation from the noisy speech distribution to the clean speech distribution is modeled by an ODE without considering the random Wiener process and is given by
\begin{equation}\label{eq:ode}
	\frac{d x_{t}}{dt}=u\left(x_{0}, t,x_{1}\right),
\end{equation}
where $u\left(x_{0}, t,x_{1}\right)$ is the vector field that governs the dynamics of the perturbed signal $x_{t}$, i.e., the flow, along the probability path. Considering a special case of an ODE whose marginal probability path $p_{t}\left(x_{t}|\mu_{t}\left(x_{1}, x_{0}\right),\sigma_{t}^{2}\right)$ is Gaussian, the vector fields can be derived based on \cite{lipmanFlowMatchingGenerative2023} as
\begin{equation}\label{eq: vetor field}
	u\left(x_{0}, t, x_{1}\right)=\frac{\sigma_{t}^{\prime}}{\sigma_{t}}\left(x_{t}-\mu_{t}\left(x_{1}, x_{0}\right)\right)+\mu_{t}^{\prime}\left(x_{1}, x_{0}\right),
\end{equation}
where $\sigma_{t}^{\prime}=\frac{d}{d t}\sigma_{t}$ and $\mu_{t}^{\prime}\left(x_{1}, x_{0}\right)=\frac{d}{d t}\mu_{t}\left(x_{1}, x_{0}\right)$ denote the time derivative of $\sigma_{t}$ and $\mu_{t}\left(x_{1}, x_{0}\right)$, respectively. Note that the standard deviation $\sigma_{t}$ is constrained to be non-zero. Like the score function, the vector field is intractable owing to the ignorance of the prior knowledge of clean speech, and thus a neural network is employed to estimate the vector field using the following flow matching loss
\begin{equation}\label{eq:fm_loss}
	\mathcal{L}_{f m}(\theta)=\mathrm{E}_{x_{t}\mid x_{1},t,x_{0}} \left[\left|u_{\theta}\left(x_{t}, t,x_{1}\right)-u\left(x_{0}, t,x_{1}\right)\right|^{2}\right],
\end{equation}
which measures the mean square error (MSE) between estimated and true vector field. Once the vector field network is trained, we can obtain the clean speech signal $x_0$ from $x_1$ based on the ODE solver for the Eq. (\ref{eq:ode}). 

In contrast to score-matching methods where perturbed signals are constrained by predefined SDEs, flow-matching methods achieve greater flexibility by directly specifying mean and variance schedules through vector-field estimation. This fundamental difference in formulation allows flow-matching approaches to bypass the architectural constraints inherent in score-based methods.
\begin{figure*}[htbp]
	\centering
	
	\subfloat[Comparison of different mean schedules]{
		\includegraphics[width=0.45\textwidth]{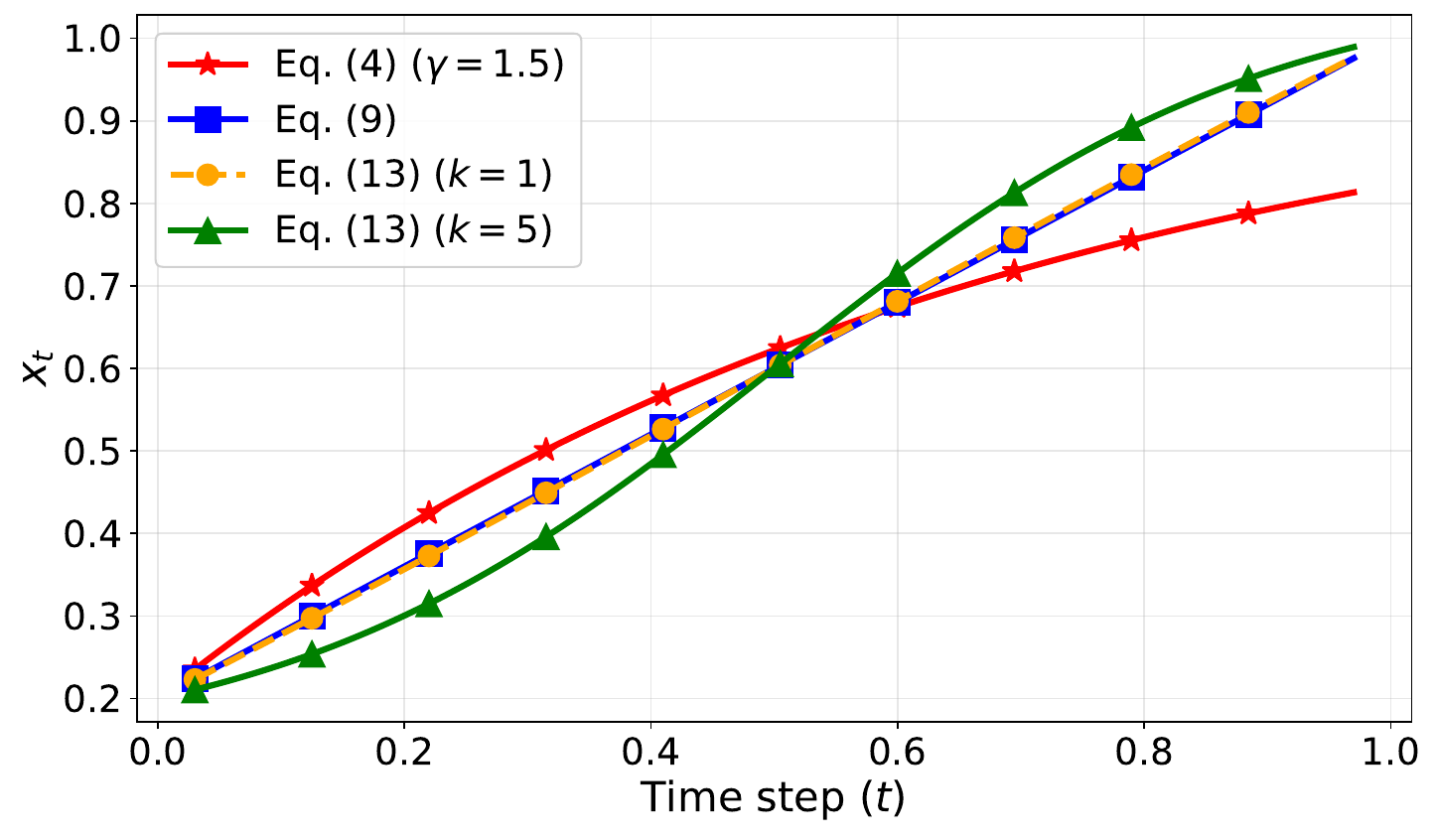}
		\label{fig:mean_schedule}
	}
	\hfill
	\subfloat[Comparison of SNR curves of different mean schedules]{
		\includegraphics[width=0.45\textwidth]{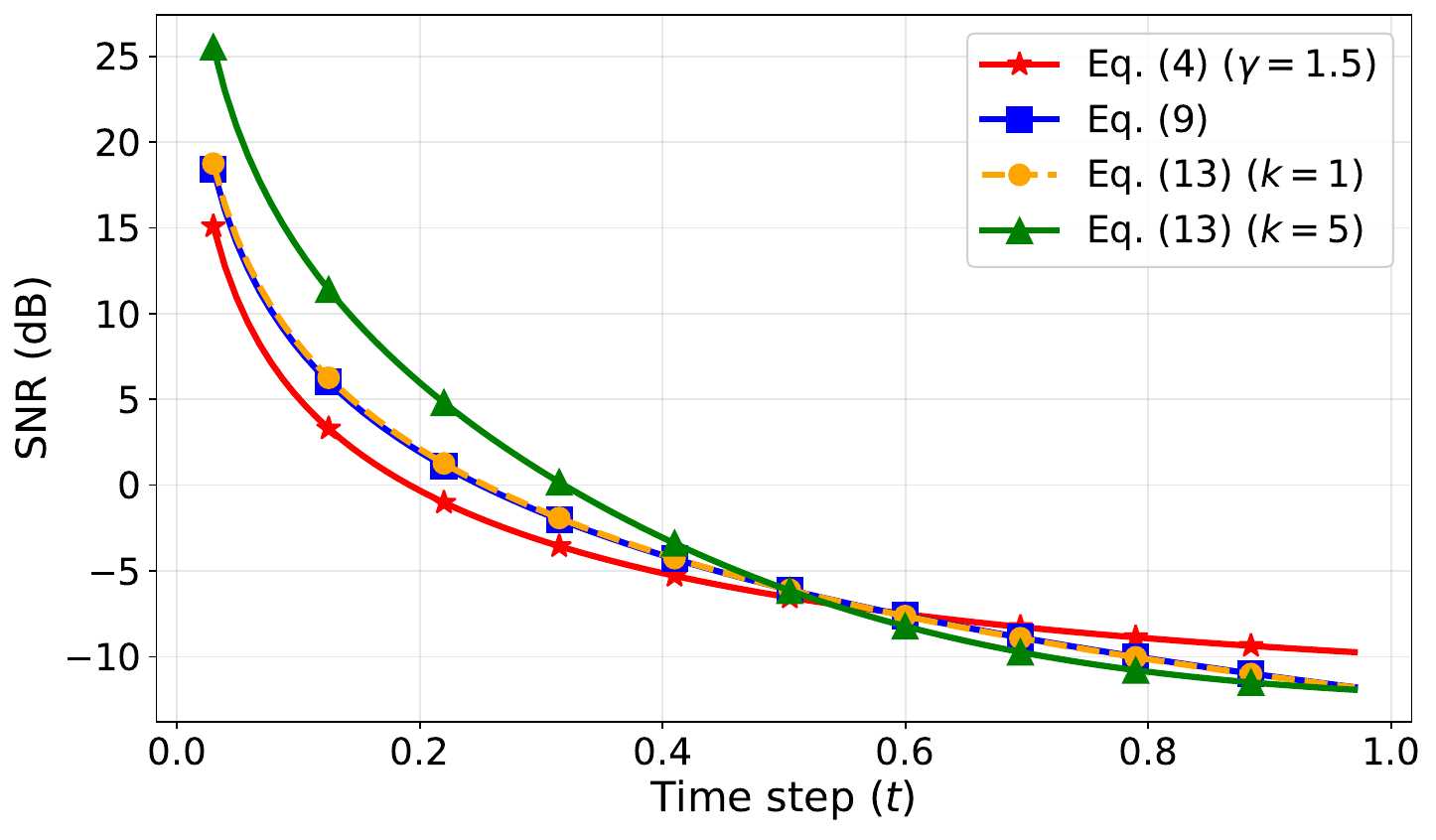}
		\label{fig:snr_curve}
	}
	
	\caption{Comparison of different mean schedules and their SNR curves. Note that $x_0$ and $x_1$ are set to 0.2 and 1, respectively.}
	\label{fig:schedules}
	
\end{figure*}

\begin{figure*}[htbp]
	\centering
	\includegraphics[width=1\textwidth]{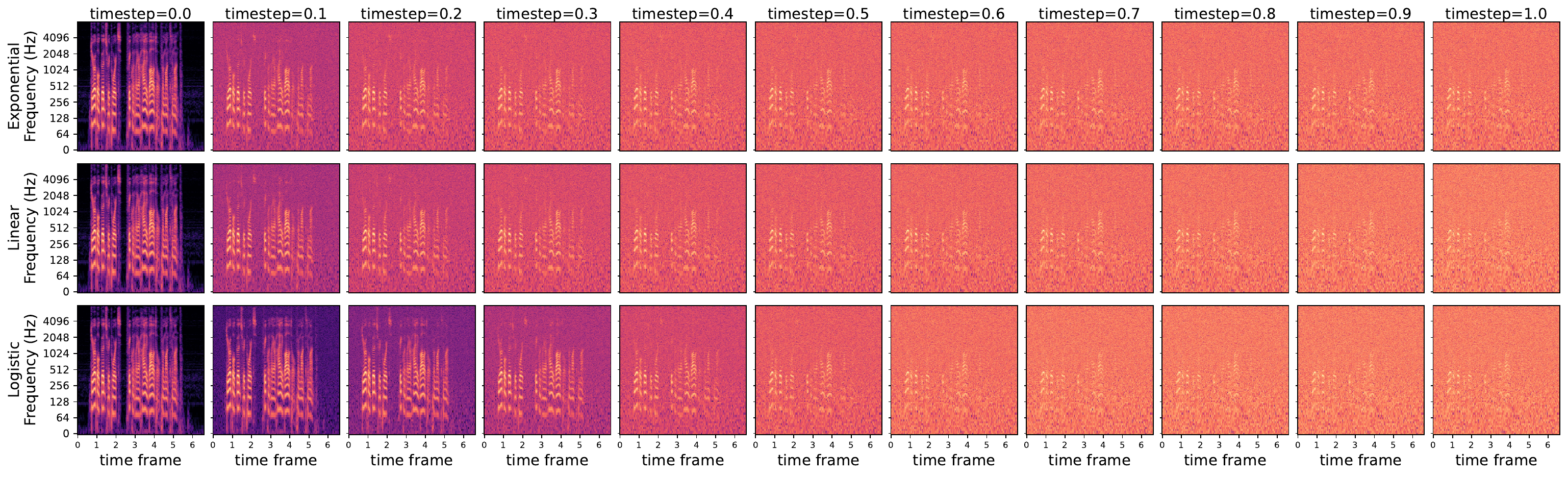}
	\caption{Comparison of spectrograms transport using different mean schedules}
	\label{fig: mean schedule spectrograms}
\end{figure*}

The flow-matching method is characterized by the pre-defined mean and variance schedules. The widely used mean and variance schedules are
\begin{equation}\label{eq: linear mean schedule}
	\mu_{t}\left(x_{0}, x_{1}\right)=(1-t) x_{0}+tx_{1},
\end{equation}
and
\begin{equation}\label{eq: linear variance schedule}
	\sigma_{t}=t\sigma,
\end{equation}
where $\sigma$ is a hyperparameter to determine the standard deviation of $x_{1}$. This indicate that both the mean $\mu_{t}$ and standard deviation $\sigma_{t}$ follow linear trajectories over time. Substituting this mean and variance schedules into Eq. (\ref{eq: vetor field}), we have the vector field
\begin{equation}
	u\left(x_{t}, t, x_{1}\right)=\frac{1}{t}  \left(x_t - \mu_{t}\left(x_{0}, x_{1}\right) \right)  +\left(x_{1}-x_{0}\right).
\end{equation}
 Through our analysis in Section IV.C, we find that the vector field inherently contains the stochastic component due to the time-varying variance schedule. This stochastic component results in inefficient training and inference processes. Moreover, we find that these stochastic components lead to unstable estimations in speech enhancement tasks, resulting in both residual noise retention and hallucination artifacts, where the system generates speech-like noise devoid of actual speech content.

\subsection{NCSN++ Backbone}
As mentioned above, a neural network is employed to parameterize the intractable score function or vector field. The most widely used backbone in audio-related tasks is the NCSN++, which is a U-Net-based diffusion model originally designed for image generation \cite{songSCOREBASEDGENERATIVEMODELING2021a} . 
While this approach works well for images, as 2D convolutions inherently capture spatial locality and multi-scale noise scheduling effectively models pixel dependencies, it shows significant limitations when processing audio signals.
First, NCSN++ struggles to properly disentangle time-frequency representations: its 2D convolutions cannot explicitly model the distinct characteristics of temporal (long-range dependencies, e.g., speech coherence) and spectral (harmonic structures, energy distribution) dimensions, often leading to phase misalignment or over-smoothed spectra in audio reconstruction.
Second, its computational cost is prohibitive for high-resolution audio tasks. With an extensive parameter space of approximately 65 million trainable weights, the network requires roughly 133 billion floating-point operations per second audio signal sampled at 16 kHz, making it inefficient for large-scale or real-time audio applications.

\section{The Proposed Method}
\subsection{Motivation and Strategy}
We try to investigate more efficient mean and variance schedules for the perturbed signal, i.e., non-linear mean and variance schedules, which may bring performance improvement. However, we find inherent limitations in both score-matching and flow-matching frameworks. In the score-matching framework, selecting a suitable SDE is very difficult, particularly for complex mean and variance schedules. In the flow-matching framework, the vector field is too complicated with the stochastic component and often leads to inefficient training and inference. These fundamental limitations motivate our proposed target-matching based generative model. In addition, we design a new diffusion backbone for the audio signal, which can process audio signal more efficiently than the widely used NCSN++ backbone.

\subsection{The Logistic Mean and Bridge Variance Schedules}

Like score-matching and flow-matching frameworks, we assume that the marginal distribution is Gaussian
\begin{equation}
	p(x_t) = \mathcal{N}(x_t; \mu_t (x_1,x_0), \sigma^2 (t)).
\end{equation}
We propose the mean follows the logistic schedule as
\begin{equation}\label{eq: logistic mean schedule}
	\mu_t (x_0,x_1) = x_0 + \frac{x_1 - x_0}{e^{k/2} - 1} \left( \frac{1 + e^{k/2}}{1 + e^{-k(t-0.5)}} - 1 \right),
\end{equation}
where $k$ denotes the steepness of the transition between the initial value $x_0$ and the target value $x_1$. A larger $k$ yields a steeper transition in the logistic mean schedule, causing the interpolation between $x_0$ and $x_1$ more abruptly around $t=0.5$.

We compare the logistic mean schedule in Eq. (\ref{eq: logistic mean schedule}) with two other popular mean schedules: the OUVE schedule (Eq. (\ref{eq:ouve mean schedule})) used in the score-based method and the linear schedule (Eq. (\ref{eq: linear mean schedule})) used in the flow-based method. Fig. \ref{fig:schedules} shows these three mean schedules and their SNR curves. It can be observed that the logistic mean schedule has three advantages.  Firstly, it is more generalized than the linear schedule and nearly reduce to the linear schedule when the steepness is small. Secondly, as noted in \cite{layReducingPriorMismatch2023}, the OUVE schedule exhibits a mean mismatch at the final time, whereas the logistic mean schedule has less mismatch than both the OUVE and linear mean schedule. Thirdly, the logistic mean schedule has a more linear SNR reduction curve, which is likey to brings more effective perturbation. 
This can be attributed to the expanded dynamic range in our logistic mean schedule, which achieves a significantly broader SNR trajectory compared to conventional linear and OUVE schedules. Specifically, near the initial diffusion phase around $t=0$, signals retain substantially higher SNR levels that preserve critical speech components with minimal spectral distortion. Conversely, at the terminal phase around $t=1$, the schedule incorporates lower SNR conditions .

To provide an intuitive comparison, we examine the spectrograms of these three mean schedules at different timesteps, as shown in Fig. \ref{fig: mean schedule spectrograms}. Our observations reveal that all three schedules perturb the signal by introducing increasing noise as the timestep $t$ evolves from 0 to 1. However, the exponential mean schedule (Eq. \ref{eq:ouve mean schedule}) exhibits two notable limitations. The first is that spectrograms for $t > 0.1$ predominantly demonstrate low SNR, which will over perturb the signal and leads to inefficient inference. The second is that the final perturbed signal at $t = 1$ remains insufficiently transformed. The linear schedule effectively mitigates these issues, while our proposed method demonstrates superior performance in achieving more balanced and effective perturbation throughout the entire diffusion process.

Different to the widely used variance exploding, variance preserving, and linear schedules, we use the bridge variance schedule
\begin{equation}
	\sigma_t = \sigma \sqrt{t(1-t)}
	\label{eq:bridge_variance},
\end{equation}
where $\sigma$ is a hyperparameter that controls the maximal Gaussian distribution. Note that $\sigma_t$ is constrained to be non-zero, and thus the timestep is constrained by $0 < t < 1$. Compared to the linear variance schedule Eq. (\ref{eq: linear variance schedule}) and the constant variance schedule used in \cite{jungFlowAVSEEfficientAudioVisual2024}, the bridge variance reduces to $0$ at $t=0$ and $t=1$ and thus is more suitable in the speech enhancement task, because both the clean and noisy signals are determined but not stochastic. 

%
%
%
\begin{figure*}[htbp]
	\centering
	\subfloat[Overall Architecture\label{fig:overal}]{
		\raisebox{-.5\height}{\includegraphics[width=0.56\linewidth]{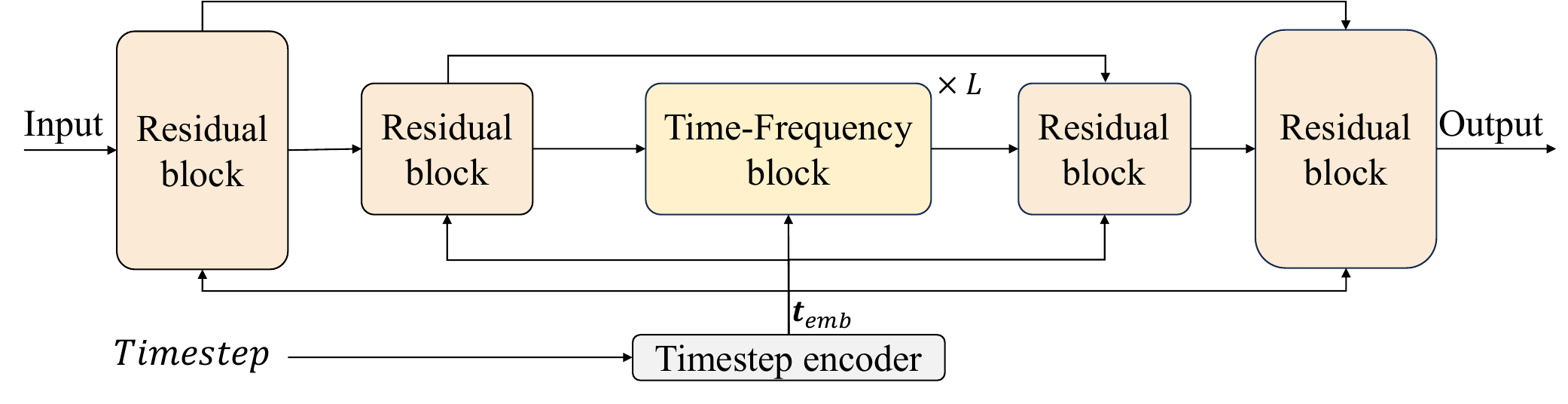}}
	}
	\hfill
	\subfloat[Residual Block Structure \label{fig:resudual block}]{
		\raisebox{-.5\height}{\includegraphics[width=0.4\linewidth]{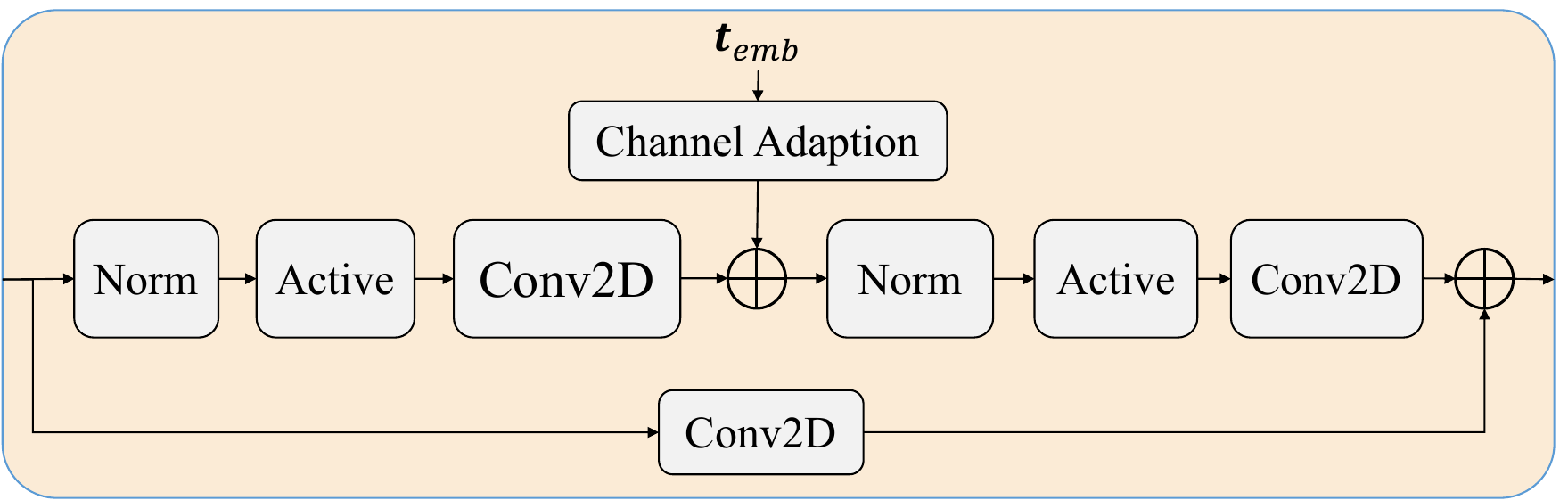}}
	}
	
	\vspace{2mm} 
	
	\subfloat[Time-Frequency Block Detail\label{fig:Time-Frequency block}]{
		\raisebox{-.5\height}{\includegraphics[width=0.37\linewidth]{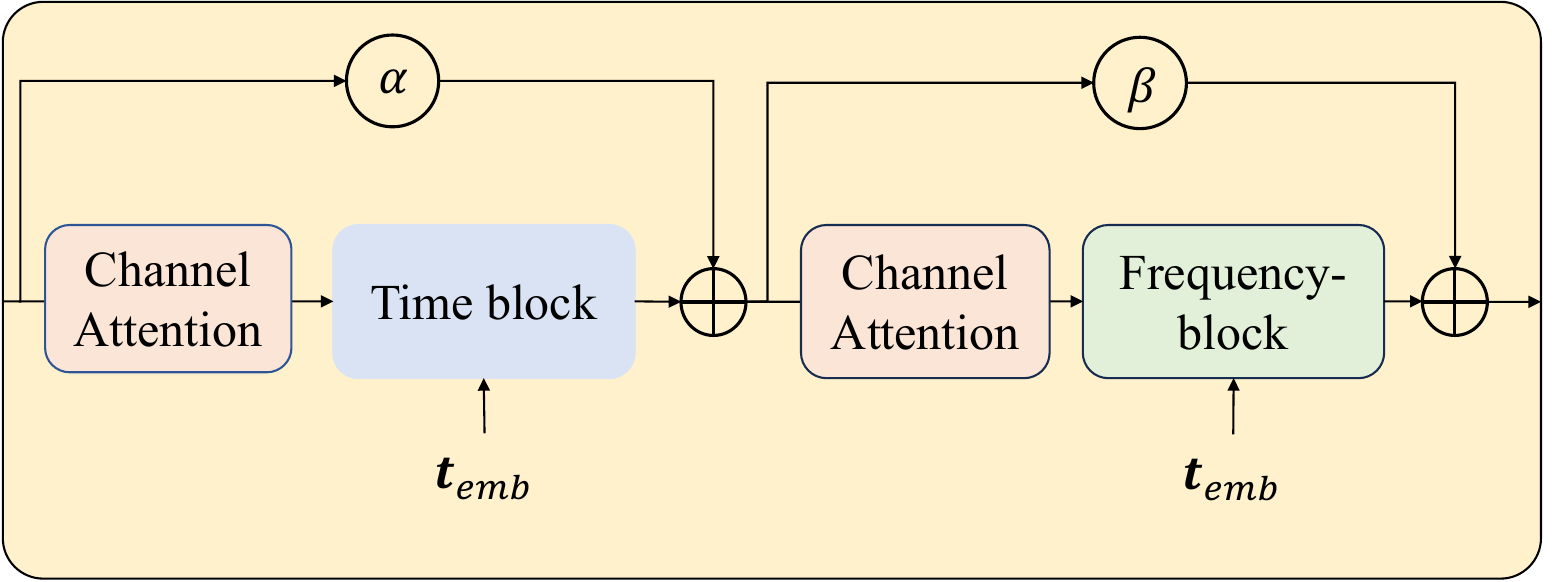}}
	}
	\hfill
	\subfloat[Time Block\label{fig:time block}]{
		\raisebox{-.5\height}{\includegraphics[width=0.31\linewidth]{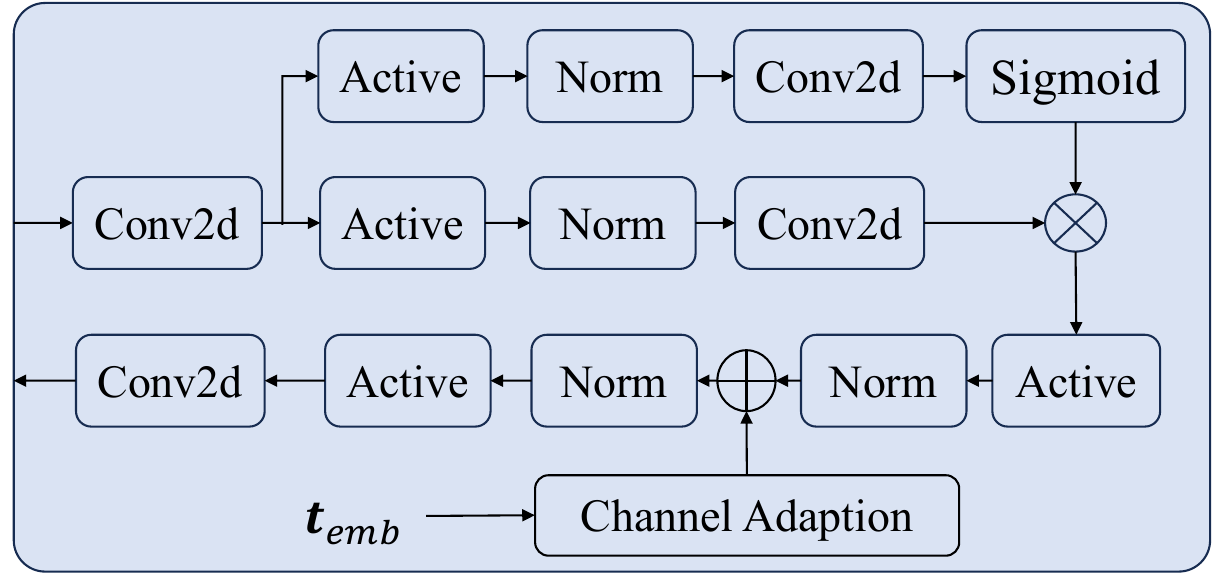}}
	}
	\hfill
	\subfloat[Frequency Block\label{fig:frequency block}]{
		\raisebox{-.5\height}{\includegraphics[width=0.275\linewidth]{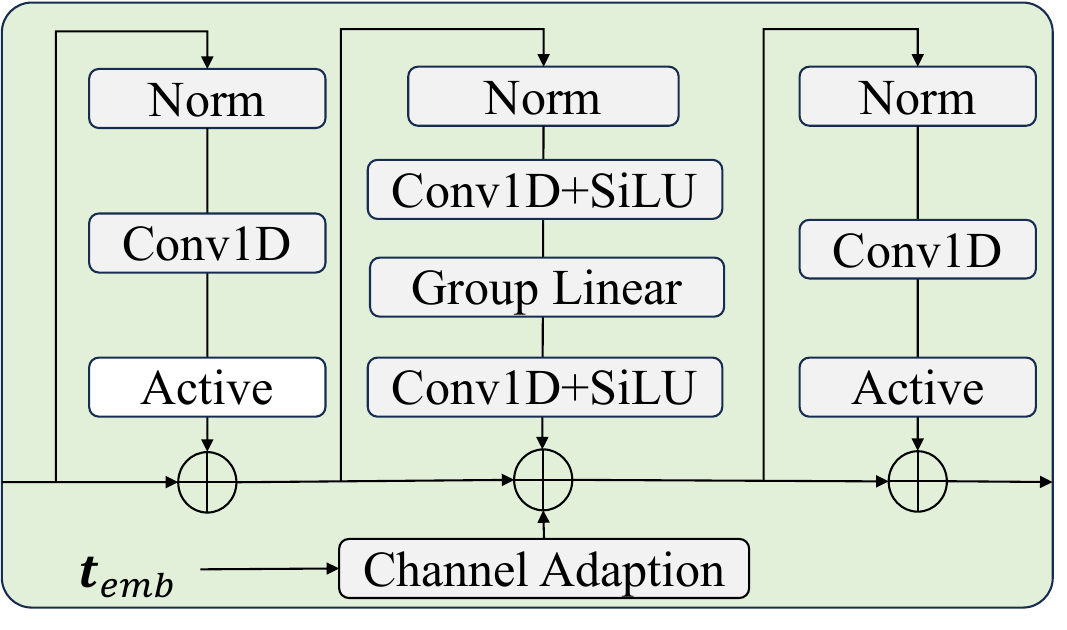}}
	}
	
	\vspace{-2mm} 
	\caption{Overall and detailed architecture of the proposed neural network, showing (a) Overall Architecture, (b) Residual block detail, (c) Time-frequency block, (d) Time block detail, and (e) Frequency block detail. The operations $\oplus$ and $\otimes$ denote the element-wise addition and  multiplication, respectively.}
	\label{fig:dba}
\end{figure*}
\subsection{Target Matching}
Based on the pre-difined mean and variance schedules, the perturbed signal can be generated using the reparameterization trick
\begin{equation}	\label{eq: generated signal}
x_t = \mu_{t}\left(x_{0}, x_{1}\right) + \sigma_tz_t,
\end{equation}
where $z_t$ is a random variable drawn from a standard Gaussian distribution at timestep $t$.
 Substituting Eq. (\ref{eq: generated signal}) into Eq. (\ref{eq: vetor field}), the vector field is given by
\begin{equation}
		u\left(x_{t}, t, x_{1}\right) = {\sigma_{t}^{\prime}} z_{t}+\mu_{t}^{\prime}\left(x_{0}, x_{1}\right).
	\label{eq:vector_field2}
\end{equation}
It can be observed that the vector field $u\left(x_{t}, t, x_{1}\right)$ is stochastic due to the existence of $z_{t}$ as long as the variance schedule is time-varying ($\sigma_t^{\prime} \neq 0 $) and thus results in inefficient training and inference processes.

To avoid this problem, we propose to match the output of the neural network to the target signal instead of the vector field, which gives the following target matching loss function
\begin{equation}
	\mathcal{L}_{tm}(\theta)=\mathbb{E}_{x_{t}\mid x_{1},t,x_{0}} \left[\left|x_{\theta}\left(x_{t}, x_{1},t\right)-x_{0}\right|^{2}\right],
	\label{eq:loss_function}
\end{equation}
which measures the MSE between estimated and true target signals. Compared to the flow-matching loss in Eq. (\ref{eq:fm_loss}), the proposed target matching loss has two advantages. The first advantage is that the predicted target does not contain any random component and hence leads to efficient training and inference processes. Second, the target matching framework enables us to incorporate signal characteristics through signal loss. Unlike vector fields, target signals possess clearer physical interpretations. In speech enhancement tasks, for instance, the target signal represents the spectral characteristics of clean speech, while vector fields typically correspond to noise components. Therefore, target signals exhibit more structured properties compared to vector fields and hence can be estimated more effectively. Furthermore, in score-based and flow-based methods, clean speech signals cannot be directly acquired but must be sampled after multiple Neural Function Evaluations (NFEs). 

We introduce two signal loss in this paper to validate the superiority of the proposed method. The first signal loss is the frequency-domain multi-scale Mel-spectrum loss
\begin{equation}
\mathcal{L}_{mel}(\theta) = \sum_{j=1}^{J} \mathcal{L}_{mel}^{j}(\theta),
\end{equation}
where $J$ denotes the scale number of Mel filterbands. 
The $j$-th Mel loss is the MSE between the estimated and true Mel-spectragrams
\begin{equation}
	\mathcal{L}_{mel}^{j}(\theta) = \frac{1}{L^{j} F_{mel}^{j}} \sum_{l,f} \left\| \hat{X}_{l,f_{mel}} ^{j}- X_{l,f_{mel}}^{j}\right\|_1,
\end{equation}
where the Mel-spectrograms $\left\{\hat{X}_{l,f_{mel}}^{j}, X_{l,f_{mel}}^{j}\right\} \in \mathbb{R}^{L^{j} \times F_{mel}^{j}}$ are calculated as $
X_{l,f_{mel}}^{j}= |X^{j}|\mathcal{A}^{j} \quad \text{and} \quad \hat{X}_{l,f_{mel}} ^{j} = |\tilde{X}^{j}|\mathcal{A}^{j}
$, and $\mathcal{A}^{j} \in \mathbb{R}^{F \times F_{mel}^{j}}$ is the $j$-th linear Mel filter.
The parameters used in computing each Mel-spectrum loss, including the STFT window size, hop size, and number of Mel bands, exhibit variability across configurations.
This multi-scale Mel-spectrum loss~\cite{NEURIPS2023_58d0e78c} mitigates information degradation inherent in dimensionality reduction and downsampling operations during Mel-spectrogram generation. By optimizing spectral representations across heterogeneous time-frequency resolutions, this approach better preserves discriminative audio attributes, yielding enhanced reconstruction fidelity and increased training stability.

The second signal loss is the time-domain scale-invariant SNR (SISNR) loss, which evaluates the waveform similarity between the estimated speech samples  $\hat{\mathbf{x}}$ and target speech samples $\mathbf{x}$ using
\begin{equation}
	\mathcal{L}_{SISNR}(\theta)= -10\log_{10}\left(\frac{\|\hat{\mathbf{x}}\|^2_2}{\|\mathbf{s}-\hat{\mathbf{x}}\|^2_2 + \epsilon}\right),
\end{equation}
where $\| \cdot \|_2$ represents the the Euclidean norm of a vector, $\epsilon=10^{-8}$ a small constant for numerical stability, and
\begin{equation}
	\hat{\mathbf{x}} = \frac{\langle \mathbf{x}, \mathbf{s} \rangle \mathbf{x}}{\|\mathbf{x}\|^2_2 + \epsilon},
\end{equation}
with $\langle \cdot, \cdot \rangle$ denoting the inner product between two vectors.
This scale-invariant metric provides robust evaluation of waveform reconstruction quality while being insensitive to amplitude scaling.

The composite loss combines the target matching loss and signal-level loss, and is formulated as
\begin{equation}
	\label{eq:final_loss}
\mathcal{L} (\theta)= \mathcal{L}_{tm}(\theta) + \lambda_1 \mathcal{L}_{mel} (\theta) + \lambda_2 	\mathcal{L}_{si-SNR}(\theta),
\end{equation}
where $\lambda_1$ and $\lambda_2$ denotes the weight hyperparameters.
\subsection{Diffusion Backbone for Audio}

In this paper, we propose a novel diffusion backbone for audio processing, named DBA, to estimate the target signal. 
Fig. \ref{fig:dba} shows the detailed architecture of the proposed DBA. 
The network input is formed by 
concatenating $x_t$ and $x_1$ along the channel dimension; each is a two-channel tensor concatenating the real and imaginary components of a complex spectrogram.
This yields an input shape of $B \times 4 \times F \times K$, where $B$ denotes the batch size, and the output is the estimated target signal $\hat{x}_0$ of size $B \times 2 \times F \times K$.

As shown in Fig. \ref{fig:overal}, the proposed backbone integrates a U-Net with $L$ time-frequency (TF) blocks, and both of them are conditoned by a timestep encoder.
The timestep encoder uses a sequential architecture: a Fourier feature mapping followed by linear layers and activation functions. It transforms a scalar timestep $t$ into a vector embedding $\bm{t}_{emb}$, which serves as a conditional input to all modules.
The U-Net is composed of residual blocks. As shown in Fig. \ref{fig:resudual block}, the adopted residual block differs from the original one used in NCSN++ in two key aspects. First, while NCSN++ relies on finite impulse response filters for downsampling and upsampling, our method adopts strided convolutions instead. Second, unlike NCSN++'s approach of simultaneously applying downsampling/upsampling operations across both time and frequency dimensions, our residual block selectively operates only on the frequency dimension while preserving the original resolution in the time dimension. This targeted modification in the frequency domain allows for more efficient processing while maintaining temporal resolution.
Prior to entering the residual block, the input signal passes through a 2D convolutional layer for feature transformation, where the channel dimension is projected from 4 to $C$. 

The TF block is designed to exploit inter-frame and inter-band correlations through a dual-path mechanism, which has demonstrated strong performance in speech enhancement and separation tasks due to its effectiveness in modeling speech time-frequency spectrograms \cite{wangTFGridNetIntegratingFull2023}. 
As shown in Fig. \ref{fig:Time-Frequency block}, within each TF block, T-block and F-block operations are applied iteratively to refine the input features. 
Compared to the widely used dual-path method, we introduce two different parts into the proposed TF block inspired by \cite{linPrimeKNetMultiscaleSpectral2025a}. 
First, we introduce weighted residual connections, where $\alpha$ and $\beta$ respectively denote the weights on T-block and F-block. This enables stable training with stacked TF blocks for deeper architectures. 
Second, we employ channel attention to explore inter-channel information interactions before processing in each T-block and F-block. 
Note that to condition the TF block on the timestep, we insert the $\boldsymbol{t}_{emb}$ both into the T-block and F-block. 
For the T-block, we use the squeezed TCN module proposed in \cite{liTwoHeadsAre2021a} where the channel is squeezed into $C'$ using the first Conv2d and unsqueezed back to $C$ using the last Conv2d. As shown in Fig. \ref{fig:time block}, we add the channel adapted time embedding to the feature. 
For the F-block, we use the cross-band block proposed in \cite{quanSpatialNetExtensivelyLearning2024} where the channel is squeezed into $C'$ before the group linear layer and unsqueezed back to the original dimension after the group linear layer. As shown in Fig. \ref{fig:frequency block}, we add the channel adapted time embedding to the feature.

\subsection{Sampling Method}
\begin{table}[]
	\label{algorithm1}
	\centering
	\renewcommand\arraystretch{1.3}
	\normalsize
	\begin{tabular}{l}
		\hline
		\textbf{Algorithm 1} Training process \\
		\hline
		\quad \textbf{Input: } Training pairs $(x_{1},x_{0})$   \\
		\quad For each epoch \textbf{do}:                            \\
		\quad \quad Sample timestep $t\sim\mathcal{U}([t_{eps},T])$                                 \\
		\quad \quad Compute $\mu_t(x_1,x_0)$ and $\sigma_t$ using Eqs. (\ref{eq: logistic mean schedule}) and (\ref{eq:bridge_variance})                                 \\
		\quad \quad Sample $z \sim \mathcal{N}(0,1)$                              \\
		\quad \quad Generate perturbed signal: $x_{t}=\mu_{t}(x_{1},x_{0})+\sigma_{t}z$                             \\
		\quad \quad Estimate $\hat{x}_{0} = x_{\theta}(x_{t},x_{1},t)$                               \\
		\quad \quad Compute loss $\mathcal{L}$                          \\
		\quad \quad Backpropagate to update $x_{\theta}$                         \\
		\quad \textbf{Output:}  Optimized target predictor $x_{\theta}(x_{t},x_{1},t)$\\
		\hline
		\vspace{-4mm}
	\end{tabular}
\end{table}

\begin{table}[]
	\label{algorithm2}
	\centering
	\renewcommand\arraystretch{1.3}
	\normalsize
	\begin{tabular}{l}
		\hline
		\textbf{Algorithm 2} Inference process \\
		\hline
		\quad \textbf{Input}: noisy signal $x_{1}$, target predictor $x_{\theta}(x_{t},x_{1},t)$, 
		\\    \qquad \qquad number of sampling steps $N$   \\
		\quad \textbf{Initialization: }  $x_{T}=x_{1}$, $t=T$, $n=1$   \\
		\quad \textbf{while}   $n \leq N$ \textbf{do:}                         \\
		\quad \quad Compute the vector field $u_{\theta}(x_{t},x_{1},t)$ using Eq. (\ref{eq:vector_field})                             \\
		\quad \quad Update signal: $x_{T-n/N}=x_{t}+u_{\theta}(x_{t},x_{1},t)/N$                            \\
		\quad \quad Estimate $x_{\theta}(x_{t},x_{1},t)$                               \\
		\quad \quad Update time: $t=T-n/N$, $n=n+1$ \\
		\quad \textbf{Output:}  ${x}_0$\\
		\hline
		\vspace{-4mm}
	\end{tabular}
\end{table}
Once the target estimator completes training, the initial point $x_{T}$ is first drawn from the noisy signal. The vector field then can be obtained using the estimated target $x_{\theta}(x_{t},x_{1},t)$ as
\begin{equation}
	\begin{aligned}
		u_{\theta}(x_{t},x_{1},t) &= \frac{\sigma_{t}^{\prime}}{\sigma_{t}}\left(x_{t}-\mu_{t}\left(x_{\theta}(x_{t},x_{1},t),x_{1}\right)\right) \\
		&\quad +\mu_{t}^{\prime}\left(x_{1},x_{\theta}(x_{t},x_{1},t)\right).
	\end{aligned}
	\label{eq:vector_field}
\end{equation}
Subsequently, we use the Euler ODE solver to iteratively recover the clean signal
\begin{align}
	x_{T-n/N} &\approx x_{t}+u_{\theta}(x_{t},x_{1},t)/N \label{eq:update_rule}, \\
	t &= T-n/N \label{eq:time_update},
\end{align}
where $N$ represents the total number of timesteps, $n$ is the current step, and $t$ is initialized as $T$ (0.97 for numerical stability in this paper). The use of Euler solver provides a straightforward and efficient method for solving the ODE, ensuring stable and reproducible sampling results.

We summarize the training and inference processes in Algorithm 1 and Algorithm 2, respectively. At training time, the DBA-parameterized target predictor $x_{\theta}(x_{t},x_{1},t)$ is optimized to approximate the clean signal given the noisy signal, timestep and perturbed signal. The timestep is uniformly sampled from the interval $[\epsilon, T]$, where $\epsilon$ is a small positive constant (typically 0.03) to avoid numerical instabilities near zero, and $T$ represents the maximum diffusion time. The perturbed signal is sampled as $	x_{t} = \mu_{t}(x_{1},x_{0}) + \sigma_{t}z
$ where $z$ is sampled from a standard normal distribution with zero mean and identity covariance. The $\mu_{t}(x_{1},x_{0})$ and $\sigma_{t}$ is calculated using the predefined mean and variance schedules, respectively. At the inference process, the $x_{1}$ is sampled as the noisy input signal, and then the estimated signal is calculated iteratively using the Euler ODE solver.

\section{Experimental Settings}
\subsection{Datasets}
We consider two widely used noisy speech datasets, namely VoiceBank-DEMAND (VB-DMD) and WSJ0-CHiME3. The VB-DMD dataset is a publicly available dataset generated by mixing clean speech from the VCTK dataset \cite{yamagishi2019cstr} with eight real-recorded noise samples from the DEMAND database \cite{thiemann2013diverse} and two artificially generated noise samples (babble and speech shape) at SNRs of 0, 5, 10, and 15 dB. The SNRs for the test set are 2.5, 7.5, 12.5, and 17.5 dB. The training dataset is split into a training and validation dataset, using speakers “p226” and “p287” for validation, as in [14], [16], [20]. The WSJ0-CHIME3 is generated using WSJ0 clean speech \cite{garofolo2007csr} and CHiME3 noise \cite{barkerThirdCHiMESpeech2015a} and the mixture SNR is sampled uniformly in [-6, 14] dB [18]. Approximately 13k utterances (25 h) are generated for the training set, 1.2k utterances (2 h) for the validation set and 650 utterances (1.5 h) for the test set.
\subsection{Experimental Setup}
Following \cite{richterSpeechEnhancementDereverberation2023}, we process all speech signals at 16 kHz by first computing complex-valued spectrograms $\mathbf{Y}_c \in \mathbb{C}^{256\times256}$ using STFT with the window size of 510 samples, hopsize of 128 samples, and segments of 256 frames, then apply element-wise magnitude compression $\mathbf{Y} = \beta|\mathbf{Y}_c|^\alpha e^{j\angle\mathbf{Y}_c}$ with $\alpha=0.5$ and $\beta=0.33$, subsequently stacking real/imaginary parts to form two-channel representations $\mathbf{Y} \in \mathbb{R}^{2\times256\times256}$. The neural network $x_\theta(x_t,x_1,t)$ is trained for 200 epochs using Adam optimizer ($\eta=10^{-4}$, batch size 8) with 0.999 exponential moving average on parameters, where enhanced signals are obtained by inverting the spectrogram transformations while preserving time-domain characteristics through the reversible processing pipeline.

We use only the target matching loss in the following experiments unless otherwise noted. When the composite loss is used, the weight hyperparameters $\lambda_1$ and $\lambda_2$ are set to 0.1 and 0.01, respectively. 
The multi-scale Mel loss configuration follows \cite{lei2025bridgevoc} with $J=6$, where the frame sizes of the STFT are set to $\{32, 64, 128, 256, 512, 1024, 2048\}$ and the corresponding numbers of Mel frequency bins are $\{5, 10, 20, 40, 80, 160, 210\}$. For all scales, the maximum frequency is fixed at the Nyquist frequency (half the sampling rate). The number of fast Fourier transform points is set to the frame size, and the hop size is defined as one-quarter of the frame size.

The NFE is set to 4 in the Euler solver (\ref{eq:update_rule}), unless otherwise noted.

\begin{table}
	\captionsetup{justification=centering, labelsep=newline}
	\centering
	\renewcommand{\arraystretch}{1.3}
	\caption{Architectural Configurations and Computational Complexity of DBA Variants vs. NCSN++}
	\label{tab:architecture}
	\begin{tabular}{lcccccc}
		\toprule
		Models & $C$ & $C'$ & $L$ & Param (M) & Flops (G) \\
		\midrule
		NCSN++ & 64 & -- & -- & 66 & 133 \\
		DBA-S & 32 & 96 & 4 & 3.7 & 30 \\
		DBA-M & 64 & 128 & 4 & 10.44 & 88.69 \\
		DBA-L & 96 & 192 & 4 & 23.47 & 199.31 \\
		\bottomrule
	\end{tabular}
\end{table}
\begin{table*}
	\captionsetup{justification=centering, labelsep=newline}
	\centering
	\renewcommand{\arraystretch}{1.3}
	\caption{Speech Enhancement Metrics (Mean $\pm$ Standard Deviation) of Different Backbones on WSJ0-CHIME3 dataset}
	\label{tab:dba_performance}
	\begin{tabular}{lccccccc}
		\toprule
		\textbf{Models} & \textbf{PESQ} & \textbf{SI-SDR} & \textbf{ESTOI} & \textbf{SIG} & \textbf{BAK} & \textbf{OVL} & \textbf{DNSMOS} \\
		\midrule
		NCSN++ & $2.61 \pm 0.60$ & $16.15 \pm 4.21$ & $0.89 \pm 0.08$ & $2.96 \pm 0.35$ & $3.46 \pm 0.54$ & $2.42 \pm 0.30$ & $3.90 \pm 0.24$ \\
		DBA-S  & $2.72 \pm 0.57$ & $16.24 \pm 3.97$ & $0.90 \pm 0.07$ & $2.95 \pm 0.35$ & $3.50 \pm 0.53$ & $2.43 \pm 0.29$ & $3.91 \pm 0.24$ \\
		DBA-M  & $2.81 \pm 0.57$ & $16.79 \pm 3.95$ & $0.91 \pm 0.07$ & $2.95 \pm 0.35$ & $3.49 \pm 0.53$ & $2.43 \pm 0.30$ & $3.92 \pm 0.23$ \\
		DBA-L  & $2.86 \pm 0.56$ & $17.08 \pm 3.89$ & $0.91 \pm 0.06$ & $2.94 \pm 0.35$ & $3.51 \pm 0.51$ & $2.43 \pm 0.29$ & $3.93 \pm 0.23$ \\
		\bottomrule
	\end{tabular}
\end{table*}
\subsection{Baseline Models}
We compare the proposed generative model with seven baseline models. 

SGMSE \cite{richterSpeechEnhancementDereverberation2023}: A score-based generative model for speech enhancement. 

StoRM \cite{lemercierStoRMDiffusionBasedStochastic2023}: A hybrid stochastic regeneration model consists of a predictive NCSN++ module and a diffusion-based SGMSE+ module.

SBVP \cite{jukicSchrodingerBridgeGenerative2024}: A speech enhancement model based on the Schrödinger bridge with the variance preserving noise schedule.

SBVE \cite{jukicSchrodingerBridgeGenerative2024}: A speech enhancement model based on the Schrödinger bridge with the variance exploding noise schedule.

CFM \cite{leeFlowSEFlowMatchingbased2025}: A speech enhancement model based on conditional flow matching, where the mean and the covariance for the conditional probability path are configured to move linearly from noisy speech to clean speech and decrease linearly, respectively.

Modified FM \cite{korostikModifyingFlowMatching2025}: A flow matching method for speech enhancement, with several training- and inference-time modifications.  

For all above baseline models, we used their official results reported in the original paper.

\subsection{Evaluation Metrics}
To assess the performance of speech enhancement methods, we use both the intrusive and non-intrusive metrics. For the instrusive metric, the  wideband perceptual evaluation of speech quality (PESQ) scores \cite{862PerceptualEvaluation2001}, extended short-time objective intelligibility (ESTOI) \cite{jensenAlgorithmPredictingIntelligibility2016a} and time-domain scale-invariant signal-to-distortion ratio (SI-SDR) \cite{rouxSDRHalfbakedWell2019a} are used. For the non-intrusive metrics, the DNSMOS P. 808 \cite{reddy2021dnsmos}, and DNSMOS P.835 \cite{reddy2022dnsmos} including SIG, BAK, and OVRL are used. Higher values for these metrics indicate a better speech enhancement performance.

\section{Results and Discussions}
\subsection{Investigation on the Backbone}

In this experiment, we investigate the effectiveness of the proposed DBA. Table \ref{tab:architecture} compares the architectural configuration, parameter counts and computational complexity between the proposed DBA variants (DBA-S, DBA-M, and DBA-L) and the widely adopted NCSN++ baseline. The DBA architecture introduces several modifications, including a reduced number of feature channels ($C$), squeezed dimensions ($C'$), and the number of TF blocks ($L$), resulting in significantly improved parameter efficiency. Specifically, the DBA-M model with $C=64$ achieves comparable performance to NCSN++ ($C=64$) while reducing the parameters by 84.2$\%$ (from 66M to 10.44M) and computational cost by 33.3$\%$ (from 133G to 88.69G Flops). The compact DBA-S variant further demonstrates the scalability of our architecture, achieving even greater efficiency with only 3.7M parameters and 30G Flops.

Table \ref{tab:dba_performance} presents the speech enhancement performance on the WSJ0-CHiME3 dataset, evaluated using multiple metrics. The results demonstrate that all DBA variants outperform the NCSN++ baseline across all evaluation metrics. Notably, all the three DBA variants achieves the better performance than the NCSN++ baseline. This superior performance, coupled with the substantial reduction in computational complexity shown in Table 1, validates the effectiveness and efficiency of the proposed DBA architecture for speech enhancement tasks.
The consistent performance improvement across all metrics, particularly for the DBA configuration, suggests that our architectural modifications successfully enhance the model’s capability to recover speech signals while maintaining computational efficiency. The scalability of the DBA framework is further evidenced by the progressive performance improvement from DBA-S to DBA-L, demonstrating the architecture's adaptability to different computational budgets. The standard DBA-M is used for the following experiments.
\begin{figure*}[htbp]
	\centering
	\subfloat[PESQ]{
		\includegraphics[width=0.46\textwidth]{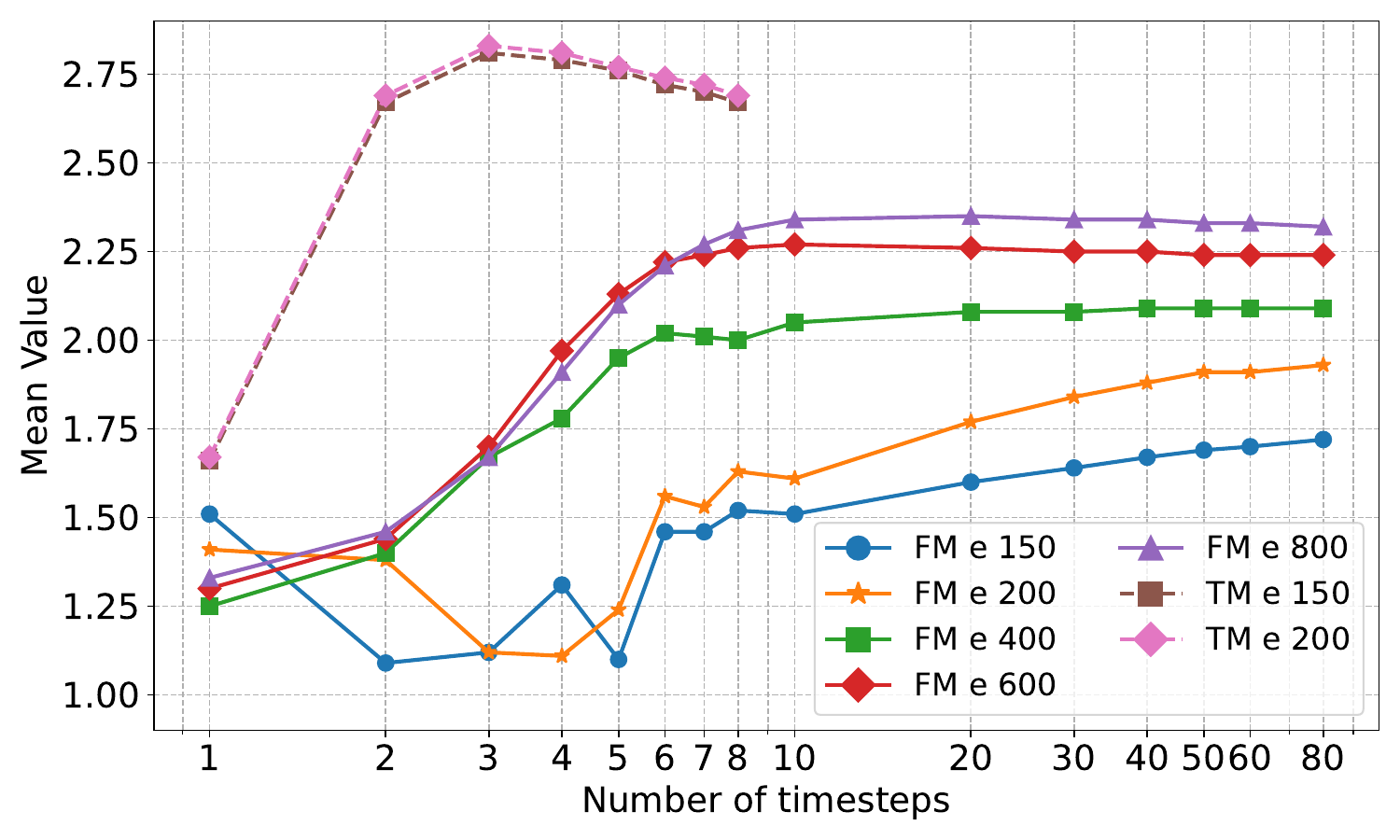}
		\label{fig:PESQ}
	}\hfill
	\subfloat[SI-SDR]{
		\includegraphics[width=0.46\textwidth]{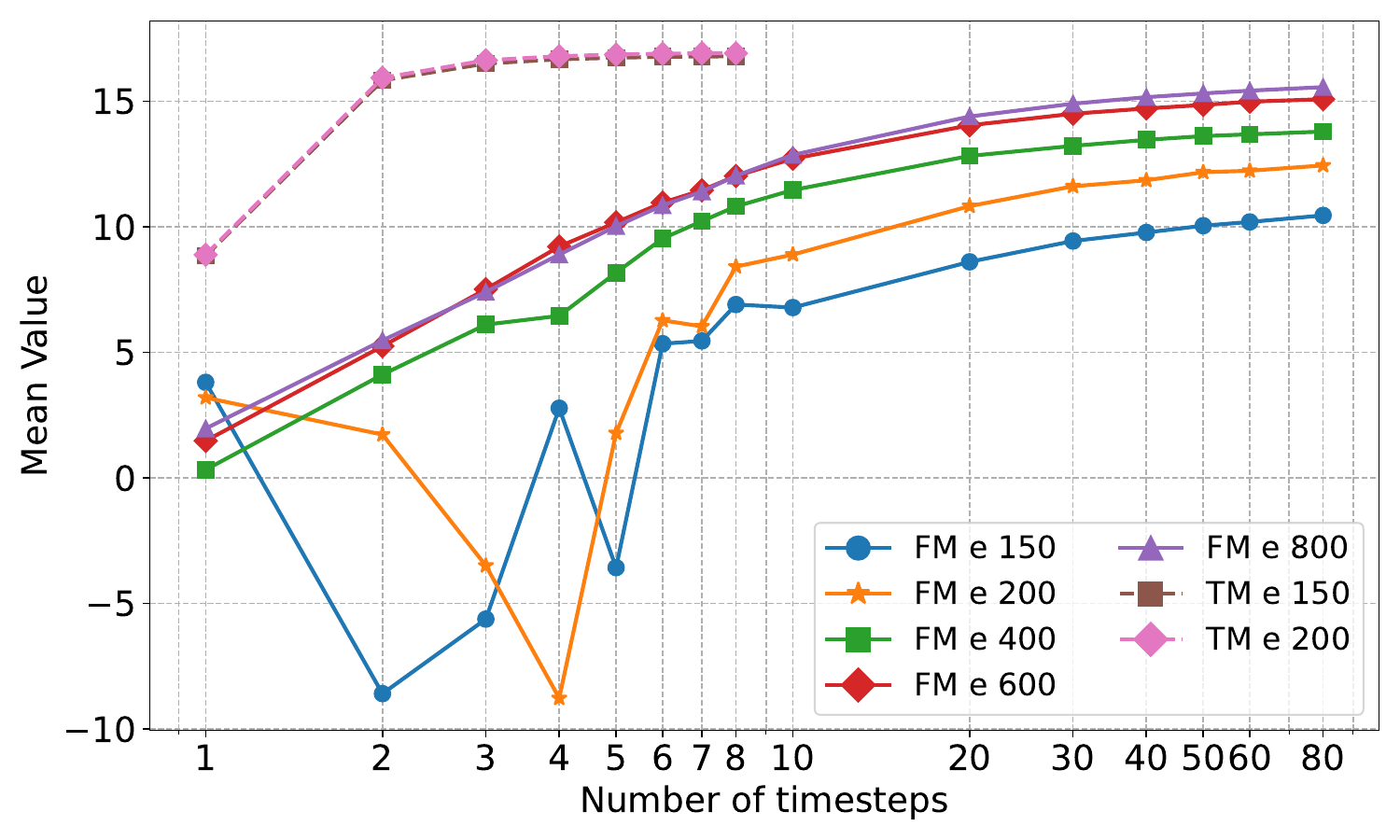}
		\label{fig:SI-SDR}
	}\hfill
	\subfloat[ESTOI]{
		\includegraphics[width=0.46\textwidth]{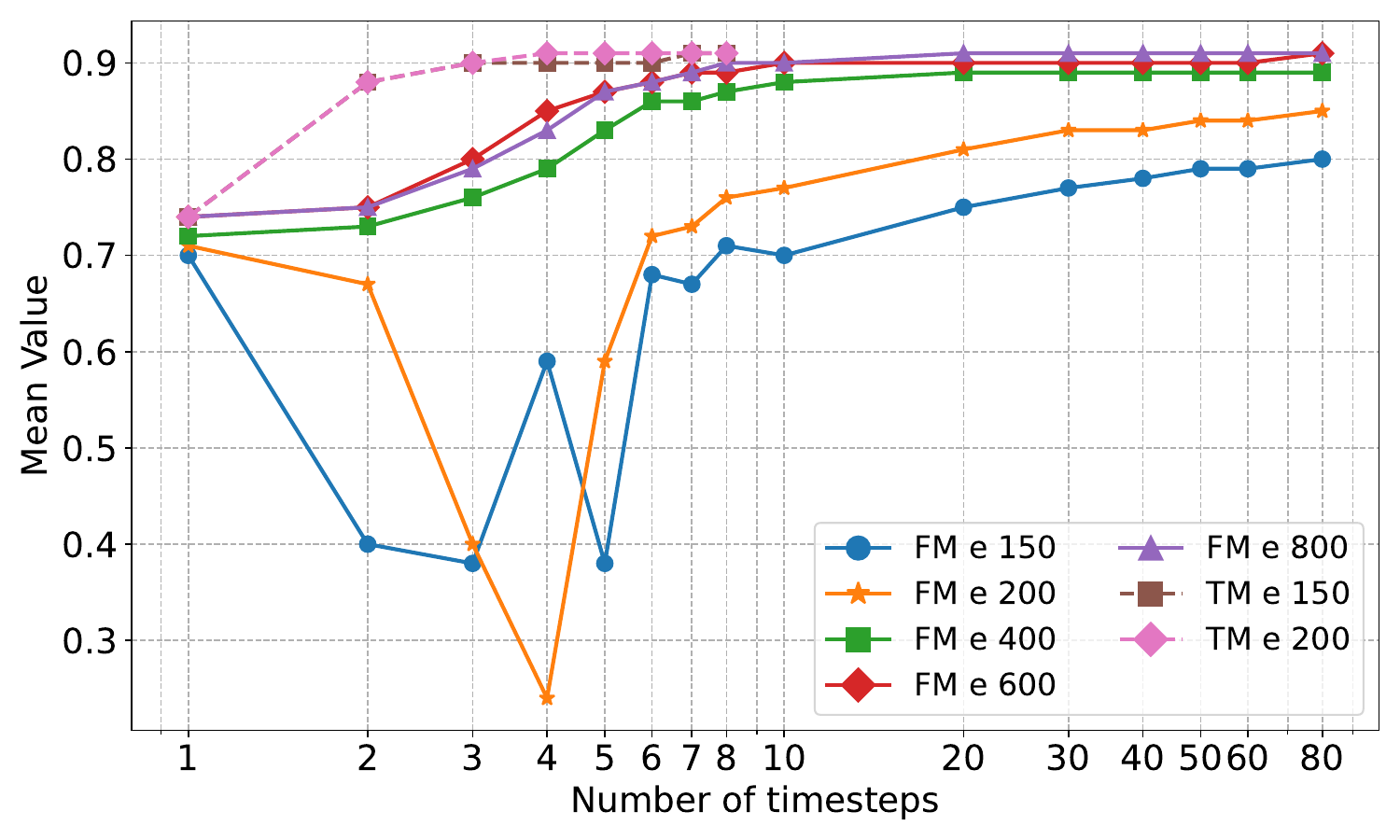}
		\label{fig:ESTOI}
	}\hfill
	\subfloat[DNSMOS]{
		\includegraphics[width=0.46\textwidth]{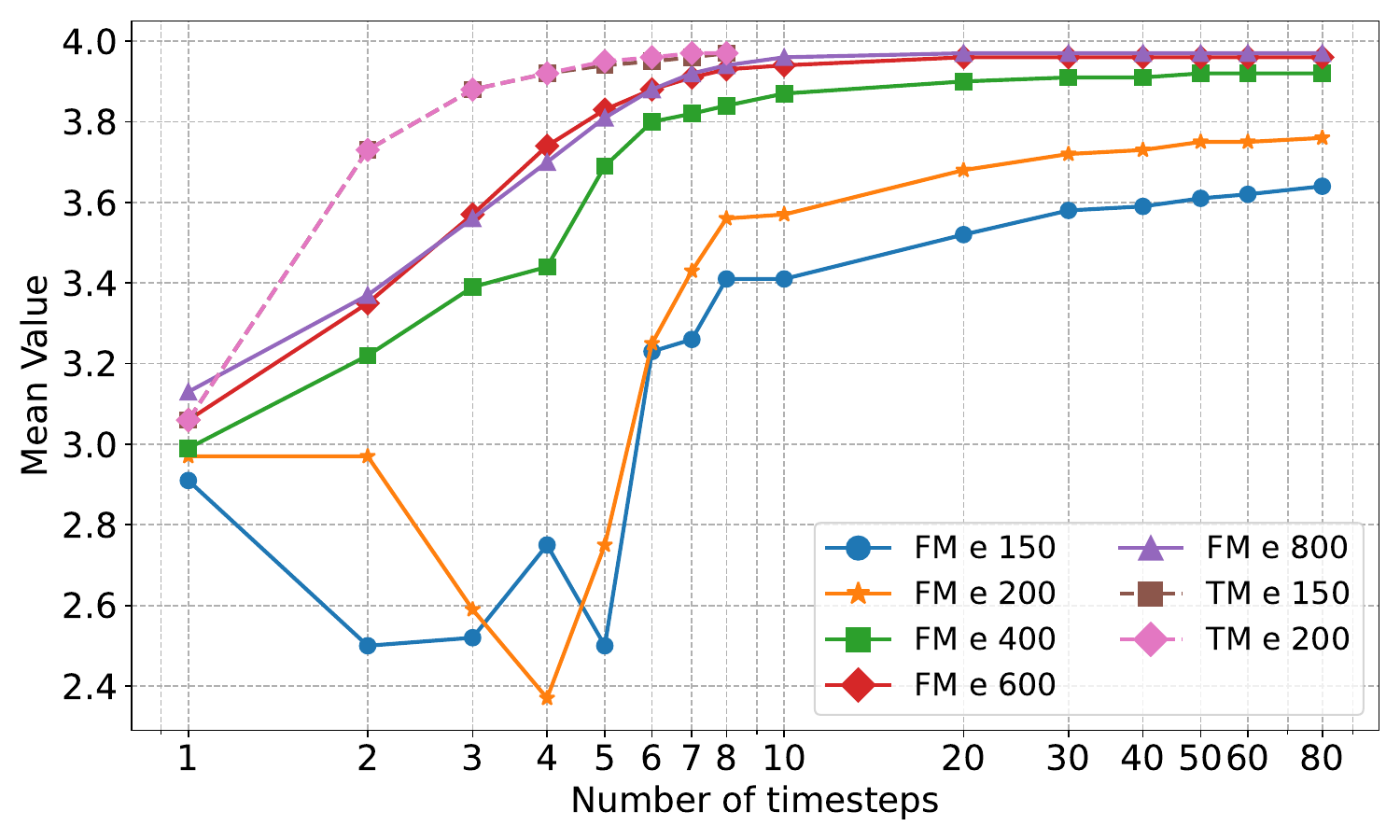}
		\label{fig:DNSMOS}
	}
	\caption{Comparison of training and inference convergence on difference metrics, best viewed in color. The `e' in the legend denotes epoch.}
	\label{fig:convergence}
\end{figure*}

\subsection{Investigation on the Noise Schedule}

To validate the effectiveness of our proposed mean and variance scheduling methods, we compare three distinct configurations as detailed in the Table \ref{table: noise schedule}.
The first approach (Schedule 1) employs the conventional mean and variance trajectories used in the traditional flow-matching based methods.
The third approach (Schedule 3) adopts our proposed mean and variance scheduling trajectories.
The second approach (Schedule 2) combines the conventional mean trajectory with our proposed variance schedule.
Experimental results for the speech enhancement on the WSJ0-CHiME3 dataset under these scheduling methods are presented in the Table \ref{tab:schedule_metrics}. It can be observed that the noise schedule has a significant impact on the enhancement performance. 
Compared to the traditional method (Schedule 1), incorporating our bridged variance schedule alone yields significant improvements in PESQ and SI-SDR, albeit at a marginal cost to ESTOI performance.
Critically, augmenting this with our logistic mean schedule not only mitigates the ESTOI degradation but also delivers comprehensive improvements across all metrics. These results demonstrate the effectiveness of the proposed mean and variance schedules. 

\begin{table}
	\captionsetup{justification=centering, labelsep=newline}
	\caption{Difference noise schedule}
	\setlength{\tabcolsep}{4pt}
	\renewcommand{\arraystretch}{1.3}
	\centering
	\label{table: noise schedule}
	\arrayrulecolor{black}
	\renewcommand{\arraystretch}{1.8} 
	\begin{tabular}{lcc}
		\arrayrulecolor{black}\hline
		Method & Mean schedule $\mu_{t}(x_{1},x_{0})$  & Variance schedule $\sigma_{t}$  \\
		\arrayrulecolor{black}\hline
		Schedule 1 & $tx_{0}+(1-t)x_{1}$ & $\sigma t$ \\
		\hline
		Schedule 2 & $tx_{0}+(1-t)x_{1}$ & $\sigma\sqrt{t(1-t)}$ \\
		\hline
		Schedule 3 & $x_{0}+\frac{x_{1}-x_{0}}{e^{k/2}-1}\left(\frac{1+e^{k/2}}{1+e^{-k(t-0.5)}}-1\right)$ & $\sigma\sqrt{t(1-t)}$ \\
		\arrayrulecolor{black}\hline
	\end{tabular}
	\arrayrulecolor{black}
\end{table}


\begin{table}[htbp]
	\captionsetup{justification=centering, labelsep=newline}
	\centering
	\caption{Speech Enhancement Metrics (Mean $\pm$ Standard Deviation) for Different Schedules}
	\label{tab:schedule_metrics}
	\renewcommand{\arraystretch}{1.3}
	\begin{tabular}{lccc}
		\toprule
		\textbf{Models} & \textbf{PESQ} & \textbf{SI-SDR} & \textbf{ESTOI} \\
		\midrule
		Schedule 1 & $2.58 \pm 0.59$ & $15.66 \pm 4.04$ & $0.90 \pm 0.07$  \\
		Schedule 2 & $2.76 \pm 0.55$ & $16.26 \pm 4.06$ & $0.89 \pm 0.08$  \\
		Schedule 3 & $2.80 \pm 0.57$ & $16.82 \pm 3.95$ & $0.91 \pm 0.07$ \\
		\bottomrule
	\end{tabular}
\end{table}

\begin{table*}[htbp]
	\captionsetup{justification=centering, labelsep=newline}
	\centering
	\renewcommand{\arraystretch}{1.3}
	\caption{The Speech Enhancement Metrics (Mean $\pm$ Standard Deviation) of the Proposed Method Using Different Signal Losses}
	\label{tab:loss_performance}
	\begin{tabular}{lccccccc}
		\toprule
		\textbf{Loss} & \textbf{PESQ} & \textbf{SI-SDR} & \textbf{ESTOI} & \textbf{SIG} & \textbf{BAK} & \textbf{OVL} & \textbf{DNSMOS} \\
		\midrule
		TM only & $2.80 \pm 0.57$ & $16.82 \pm 3.95$ & $0.91 \pm 0.07$ & $2.97 \pm 0.35$ & $3.38 \pm 0.57$ & $2.40 \pm 0.30$ & $3.93 \pm 0.23$ \\
		TM+Mel & $2.89 \pm 0.54$ & {$15.55 \pm 4.18$} & $0.91 \pm 0.07$ & $2.83 \pm 0.31$ & $3.96 \pm 0.15$ & $2.60 \pm 0.26$ & $3.89 \pm 0.23$ \\
		TM+Mel+SISNR & $2.92 \pm 0.55$ & $15.97 \pm 4.21$ & $0.91 \pm 0.06$ & $2.83 \pm 0.32$ & $3.96 \pm 0.15$ & $2.60 \pm 0.27$ & $3.91 \pm 0.23$ \\
		\bottomrule
	\end{tabular}
\end{table*}
\subsection{Investigation on the Target Matching}
In this subsection, we validate the training and inference efficiency of the proposed target matching method over the flow matching method. Fig. \ref{fig:convergence} illustrates the variation curves of different evaluation metrics versus inference steps at various training epochs for both methods. Both TM and FM employ the DBA-M network architecture with logistic and bridge scheduling strategies.

We observe from Fig. \ref{fig:convergence} that when FM is trained for 200 epochs, its performance improves with increasing inference steps but demonstrates notable instability. As the training extends to 800 epochs, the inference outcomes become more stable. In contrast, the proposed TM method achieves stable inference performance after only 200 training epochs. This comparative result effectively demonstrates the superior training efficiency of our TM approach.  
The reason lies in its deterministic optimization objective, where the loss function (Eq. (17)) directly minimizes the MSE between the estimated and true clean speech signals, thereby providing stable gradient updates. In contrast, FM optimizes a stochastic objective through its loss function (Eq. (8)), which introduces inherent noise in the vector field and leads to significant gradient fluctuations during training. 
Furthermore, it can be observed that all metrics of the FM method require more than 10 inference steps to converge, particularly the SI-SDR. In contrast, the proposed TM method typically achieves convergence within just 4 steps. This result clearly demonstrates the superior inference efficiency of our TM approach. In addition, we observed that more inference steps lead to decreased PESQ scores, which may be attributed to excessive noise suppression that compromises audio quality. This phenomenon has also been noted in other generative approaches such as FM \cite{korostikModifyingFlowMatching2025} and SBVE \cite{jukicSchrodingerBridgeGenerative2024}.

\begin{figure*}[htbp]
	\centering
	\subfloat[The spectrogram of the noisy speech]{
		\includegraphics[width=0.46\textwidth]{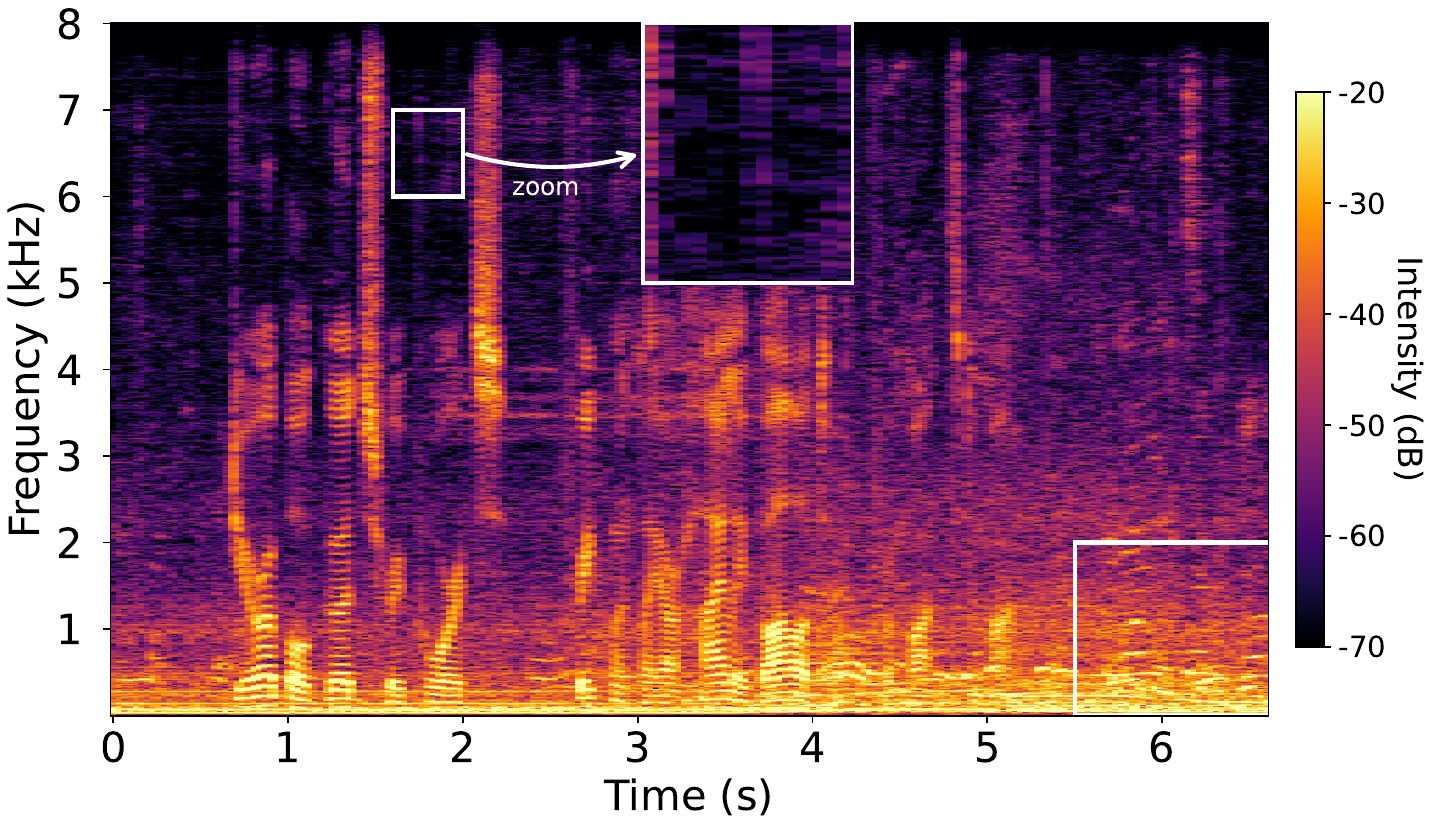}
		\label{fig:clean}
	}\hfill
	\subfloat[The spectrogram of the clean speech]{
		\includegraphics[width=0.46\textwidth]{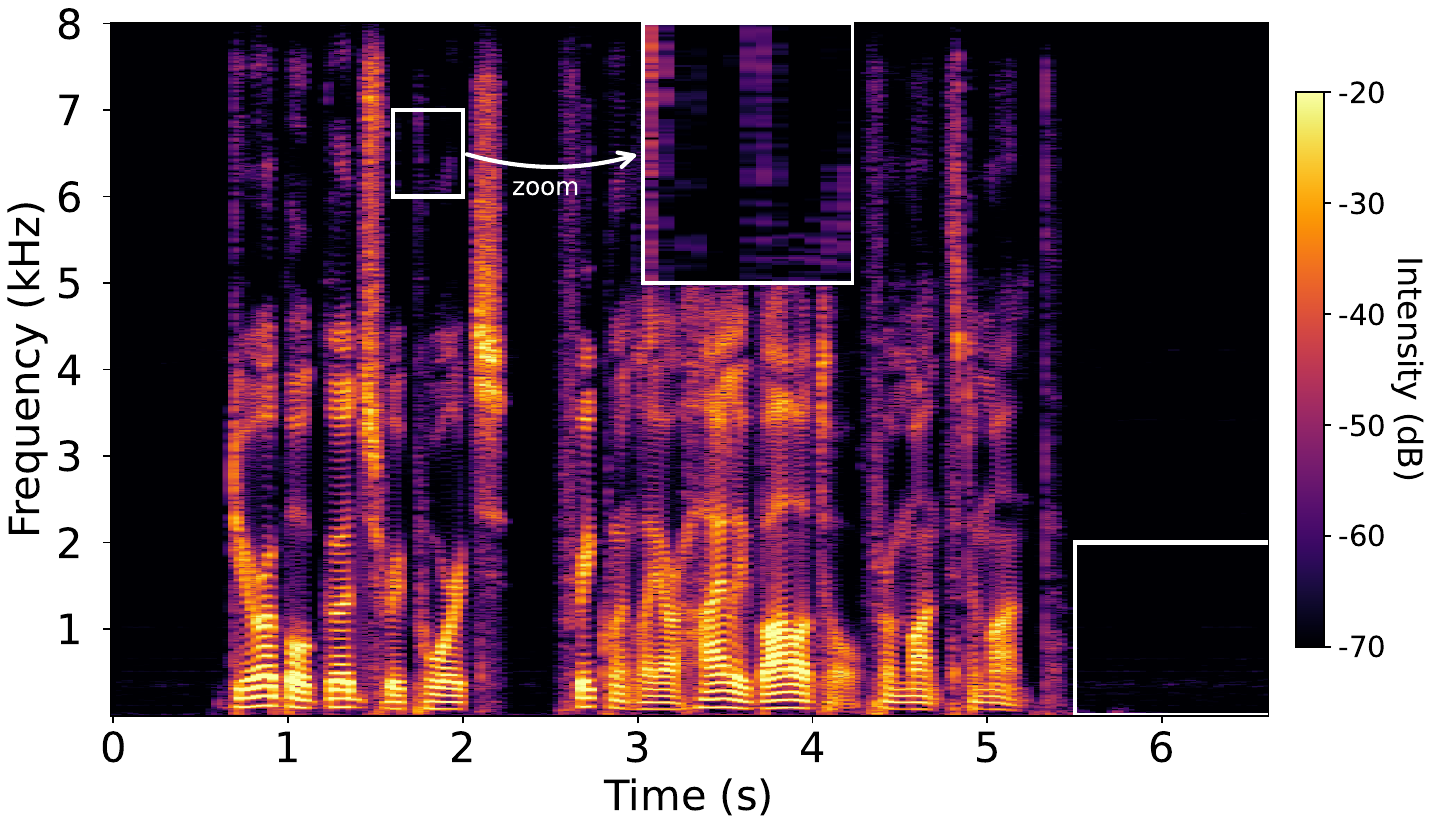}
		\label{fig:fm}
	}\hfill
	\subfloat[The spectrogram of the enhanced speech using FM]{
		\includegraphics[width=0.46\textwidth]{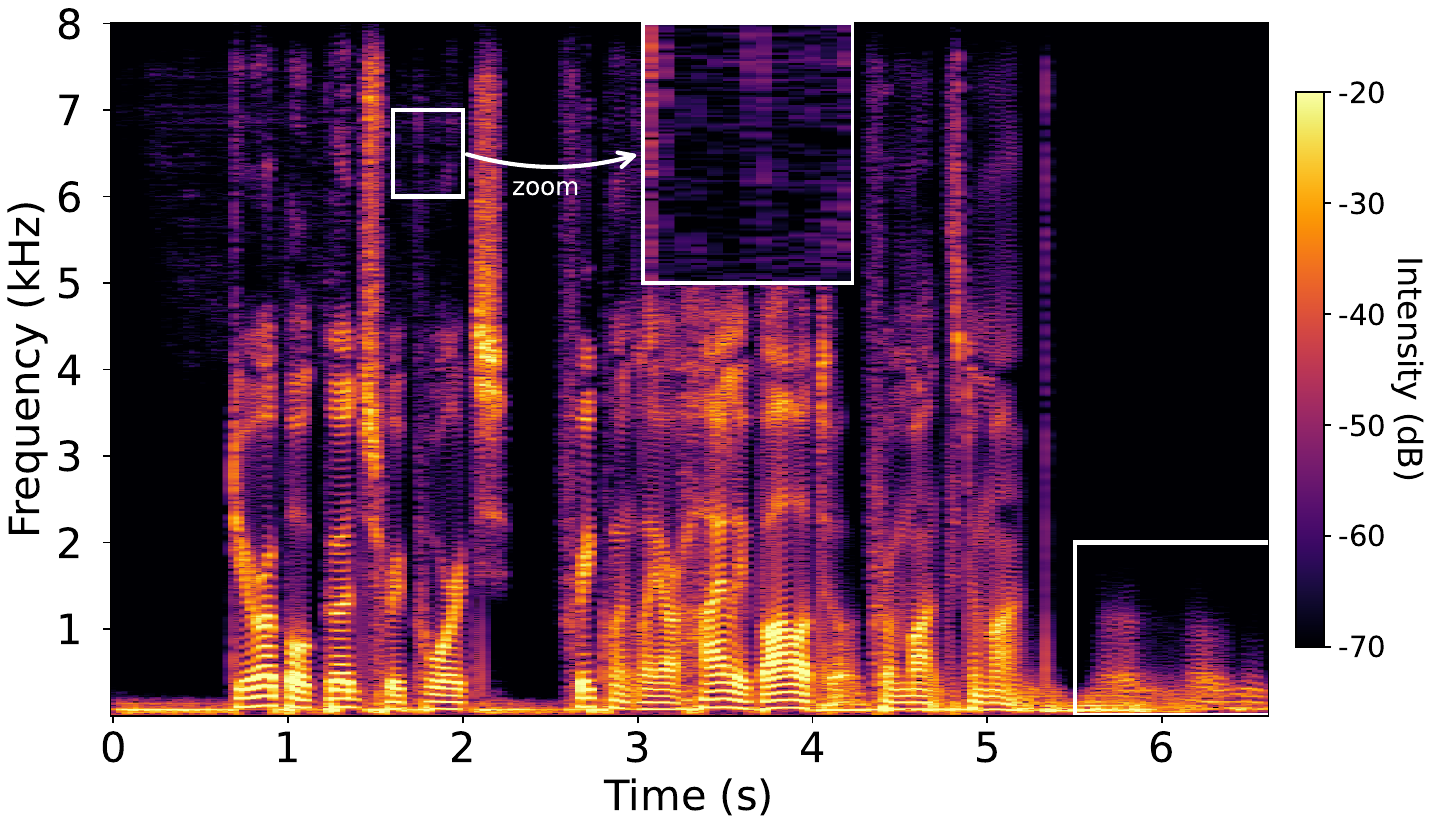}
		\label{fig:noisy}
	}\hfill
	\subfloat[The spectrogram of the enhanced speech using the proposed TM]{
		\includegraphics[width=0.46\textwidth]{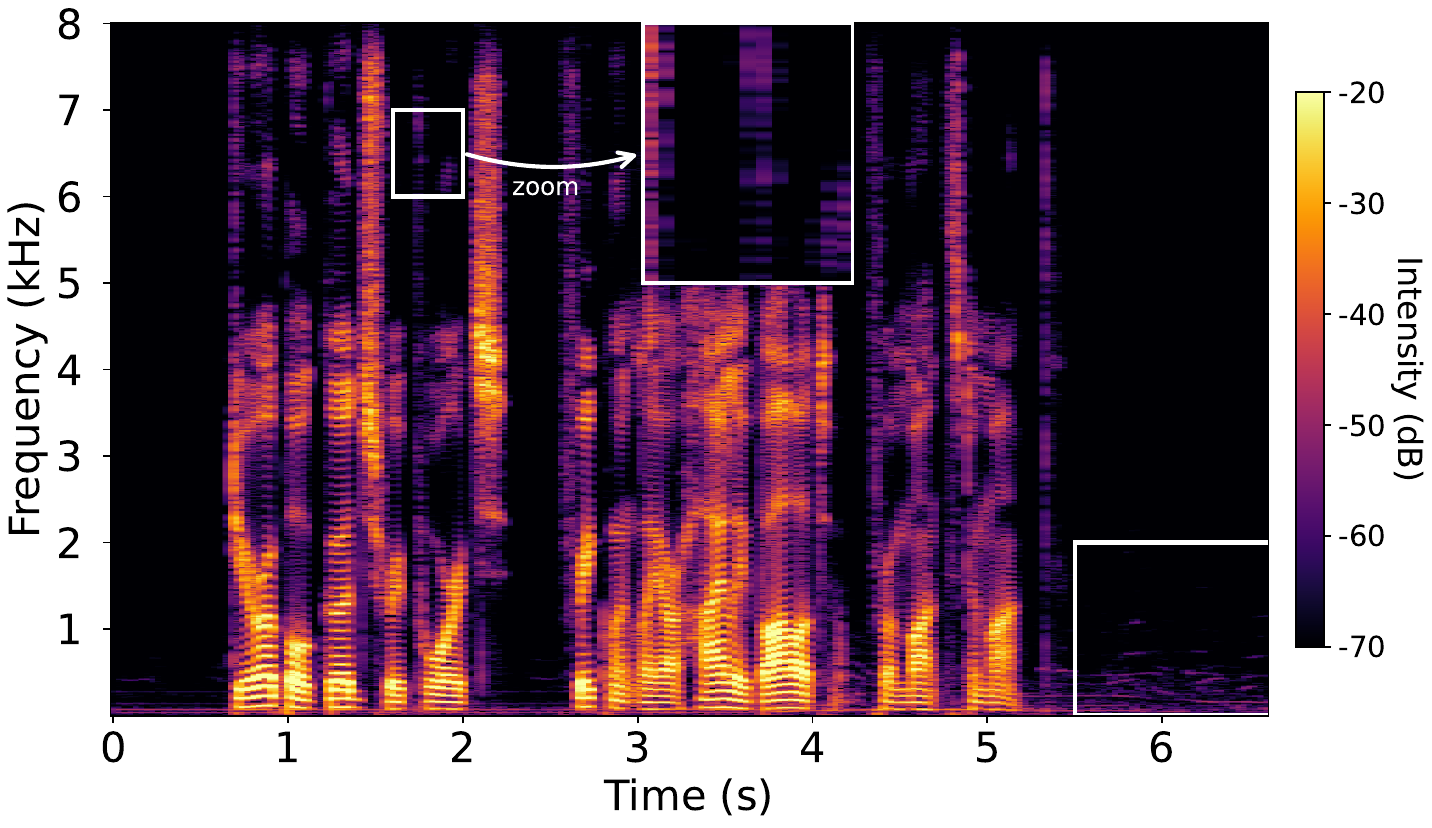}
		\label{fig:tm}
	}
	\caption{Comparison of different speech spectrogram representations.}
	\label{fig:spectrograms}
\end{figure*}

It can be observed from Fig. \ref{fig:convergence} that as the number of steps increases, the DNSMOS and ESTOI scores achieved by the FM method gradually increase and approach those obtained by the proposed TM method. Notably, however, the PESQ and SI-SDR scores of the FM method remain significantly lower than those of the proposed TM method. This persistent gap could potentially stem from the FM method's propensity to generate hallucination artifacts, which adversely affects perceptual quality metrics like PESQ and signal-level metrics like SI-SDR that rely on accurate alignment. 
To investigate this hypothesis, we present and compare the spectrograms of signals enhanced by both the FM and TM methods.
Fig. \ref{fig:spectrograms} compares the spectrograms of clean, noisy and processed audio samples by TM and FM, using the utterance "440c020w.wav" from the WSJ0-CHIME test set. By examining the rectangular regions in the left of the four subfigures, it can be observed that the FM method tends to retain noise components that resemble speech-like distributions. This behavior makes FM prone to hallucination artifacts—a common issue in generative models under low SNR conditions—whereby meaningless speech-like sounds are generated, as illustrated in the right of Fig. \ref{fig:fm}. In contrast, this phenomenon does not occur in the proposed TM-based generative approach. A plausible explanation is that, unlike FM, which relies on stochastic estimation, our method optimizes for a deterministic target, thereby reducing such artifacts.

\subsection{Investigation on the Signal Loss}
We evaluate the impact of the signal-level loss function on the performance of the proposed TM method. Experimental results are summarized in Table~\ref{tab:loss_performance}. In the baseline configuration ({TM only}), only the TM loss (Eq.~\ref{eq:loss_function}) is optimized. The second configuration ({TM+Mel}) employs the composite loss (Eq.~\ref{eq:final_loss}) with $\lambda_1=0.1$ and $\lambda_2=0$.  

It can be observed that incorporating the {Mel}-spectrum loss significantly enhances the perceptual quality and suppresses the background noise, albeit at the cost of a reduction in SI-SDR and a marginal degradation in the SIG metric. 
The integration of the SISNR loss function consistently enhances the PESQ metric, while simultaneously mitigating degradations in SI-SDR and DNSMOS induced by the TM-Mel loss function.
These results demonstrate that incorporating signal-level loss functions significantly influences the perceptual quality of enhanced audio, thereby validating a core advantage of the proposed method: its flexible integration of arbitrary signal-level loss functions, which is not feasible in score-matching or flow-matching approaches.

\begin{table}
	\captionsetup{justification=centering, labelsep=newline}
	\centering
	\renewcommand{\arraystretch}{1.3}
	\caption{Comparison Results (Mean $\pm$ Standard Deviation) of Different Methods on WSJ0-CHIME3 dataset}
	\label{tab:wsj0_chime3_results}
	\begin{tabular}{lccc}
		\toprule
		Method & PESQ & SI-SDR & ESTOI \\
		\midrule
		Noisy & $1.35 \pm 0.30$ & $4.0 \pm 5.8$ & $0.63 \pm 0.18$ \\
		SGMSE+ \cite{richterSpeechEnhancementDereverberation2023} & $2.28 \pm 0.60$ & $13.1 \pm 4.9$ & $0.85 \pm 0.11$ \\
		STORM \cite{lemercierStoRMDiffusionBasedStochastic2023} & $2.53 \pm 0.60$ & $14.8 \pm 4.3$ & $0.87 \pm 0.09$ \\
		SB-VP \cite{jukicSchrodingerBridgeGenerative2024} & $2.62 \pm 0.53$ & $14.9 \pm 4.3$ & $0.88 \pm 0.07$ \\
		SB-VE \cite{jukicSchrodingerBridgeGenerative2024} & $2.58 \pm 0.53$ & $14.7 \pm 4.2$ & $0.88 \pm 0.07$ \\
		FM & $2.32 \pm 0.66$ & $15.8 \pm 4.3$ & $0.88 \pm 0.09$ \\
		Modified FM \cite{korostikModifyingFlowMatching2025} & $2.68 \pm 0.58$ & $16.2 \pm 4.2$ & $0.89 \pm 0.08$ \\
		\midrule
		TM+DBA-M & $2.80 \pm 0.57$ & $16.8 \pm 4.0$ & $0.91 \pm 0.07$ \\
		TM+DBA-M+Signal loss & $2.92 \pm 0.55$ & $16.0 \pm 4.1$ & $0.91 \pm 0.06$ \\
		\bottomrule
	\end{tabular}
\end{table}

\begin{table}
	\captionsetup{justification=centering, labelsep=newline}
	\centering
	\caption{Comparison Results (Mean $\pm$ Standard Deviation) of Different Methods on VB-DMD dataset}
	\setlength{\tabcolsep}{4pt}  
	\renewcommand{\arraystretch}{1.3} 
	\label{tab:vb_dmd_results}
	\begin{tabular}{lcccc}
		\toprule
		Method & PESQ & SI-SDR & ESTOI & DNSMOS \\
		\midrule
		noisy & $1.97 \pm 0.75$ & $8.4 \pm 5.6$ & $0.79 \pm 0.15$ & $3.09 \pm 0.39$ \\
		OUVE \cite{richterSpeechEnhancementDereverberation2023} & $2.93 \pm 0.62$ & $17.3 \pm 3.3$ & $0.87 \pm 0.10$ & $3.56 \pm 0.28$ \\
		SBVE \cite{jukicSchrodingerBridgeGenerative2024} & $2.91 \pm 0.76$ & $19.42 \pm 3.52$ & $0.88 \pm 0.10$ & $3.59 \pm 0.30$ \\
		CFM \cite{leeFlowSEFlowMatchingbased2025} & $3.12 \pm 0.05$ & $18.95 \pm 0.23$ & $0.88 \pm 0.01$ & $3.58 \pm 0.02$ \\
								TM-DBA-M & $3.17 \pm 0.70$ & $19.44 \pm 3.35$ & $0.89 \pm 0.09$ & $3.57 \pm 0.30$ \\
		\bottomrule
	\end{tabular}
\end{table}
\subsection{Comparison Results on Speech Denoising}
We compare the proposed method against state-of-the-art generative speech enhancement models, as discussed in Section V.C. As shown in Table \ref{tab:wsj0_chime3_results}, the proposed method achieves superior performance across PESQ, SI-SDR, and ESTOI scores on the the WSJ0-CHiME3 dataset, demonstrating its effectiveness in joint perceptual quality, signal fidelity, and intelligibility optimization. Further improvement in the PESQ metric can be achieved by adopting the signal loss, but this comes at the cost of a reduction in the SI-SDR metric.

Table \ref{tab:vb_dmd_results} further compares the models on the VB-DMD dataset, where the proposed method attains the highest PESQ, SI-SDR and ESTOI scores. However, its DNSMOS is marginally lower than the optimal baseline. We note that the superiority of the proposed method is less pronounced on the VB-DMD dataset compared to the results on the WSJ0-CHiME3 dataset. This may be attributed to VB-DMD's higher average signal-to-noise ratio (SNR $>$5  dB), which aligns with our analysis in Section VI.C: the proposed method exhibits stronger hallucination suppression in low-SNR regimes, while its advantage diminishes in high-SNR conditions.

\section{Conclusion}
We have proposed a novel target matching-based generative framework for speech enhancement, addressing key limitations in existing diffusion and flow-based approaches. Our method introduces a logistic mean schedule and bridge variance schedule, which provides a more favorable SNR trajectory compared to traditional linear schedules and enables flexible and efficient training. By reformulating the problem as target matching rather than score or flow estimation, we eliminate stochastic components in the optimization objective, leading to more stable training and faster inference. Additionally, we have designed a dual-path spectro-temporal diffusion backbone explicitly optimized for audio signals, which significantly improves computational efficiency over the widely used NCSN++ architecture. Experimental results demonstrate that our framework achieves state-of-the-art performance across multiple datasets (VB-DMD and WSJ0-CHiME3) and metrics (PESQ, SI-SDR, ESTOI), while reducing model complexity by 84.2$\%$ in parameters and 33.3$\%$ in Flops compared to NCSN++.

\ifCLASSOPTIONcaptionsoff
  \newpage
\fi

\bibliographystyle{IEEEtran}
\bibliography{references}{}

\end{document}